\documentclass[11pt,oneside]{article}
\usepackage{academicons, lastpage, booktabs}
\usepackage[dvipsnames]{xcolor}
\usepackage[english]{babel}
\usepackage{amsfonts,amssymb,amsmath,amsthm}
\usepackage[absolute,overlay]{textpos}
\usepackage{float}
\usepackage{fancyhdr, graphicx}
\usepackage{enumerate}
\usepackage[margin=1in]{geometry}
\usepackage{tikz}
\usepackage{layout}
\usepackage{titlesec}
\usepackage{titling}
\usepackage{authblk}
\usepackage{xcolor, colortbl}
\usepackage{geometry}
\usepackage{siunitx} 
\usepackage{hyperref}
\usepackage{eso-pic}
\usepackage{marginnote}
\usepackage{doi}
\usepackage{wrapfig}

\def\Cref{C_{\rm ref}}

\def\Lext{L_{\rm ext}}

\def\dense{n_{\rm e}}

\def\tempe{T_{\rm e}}

\def\grho1{\left<|\nabla{\rho}|\right>}
\def\grhov1{\left<|\nabla{\rvol}|\right>}

\def\ipl{I_{\rm p}}

\def\rvol{\rho_{\rm V}}

\def\q95{q_{95}}

\providecommand{\fref}[1]{figure~\ref{#1}}

\providecommand{\Cref}[1]{Chapter~\ref{#1}}
\providecommand{\cref}[1]{chapter~\ref{#1}}

\def\Vloop{V_{\rm loop}}

\def\dense{n_{\rm e}}

\def\tempe{T_{\rm e}}

\def\Rmag{R_{\rm mag}}
\def\Zmag{Z_{\rm mag}}

\def\grho{\left \langle \left \vert \nabla\rho \right\vert\right\rangle}

\def\psiext{\psi_{\rm ext}}

\def\grho1{\left<|\nabla{\rho}|\right>}
\def\grhov1{\left<|\nabla{\rvol}|\right>}

\def\ipl{I_{\rm p}}

\def\rvol{\rho_{\rm V}}

\def\q95{q_{95}}

\def\psiplasma{\psi_{\rm pl}}
\def\psiext{\psi_{\rm ext}}
\def\psistabr{\tilde{\psi}_{\rm R}}
\def\psistabz{\tilde{\psi}_{\rm Z}}

\definecolor{coolblack}{rgb}{0.0, 0.18, 0.30}
\definecolor{carmine}{rgb}{0.59, 0.0, 0.09}
\definecolor{myblue}{rgb}{0.0, 0.48, 0.65}
\definecolor{cadetgrey}{rgb}{0.57, 0.64, 0.69}
\definecolor{lightgray}{gray}{0.9}

\hypersetup{
    colorlinks=true,
    linkcolor=coolblack,
   filecolor=magenta,      
    urlcolor=cyan,
    citecolor=coolblack,
    pdfhighlight=/P,
                 }
                                  
\usepackage{libertinus}
\usepackage[T1]{fontenc}
\usepackage{orcidlink} 

\usepackage{caption}
\usepackage{fancyhdr}

\titleformat*{\section}{\Large\bfseries\color{black}}
\titleformat*{\subsection}{\large\bfseries\color{black}}
\titleformat*{\subsubsection}{\bfseries\color{black}}

\makeatletter
\renewcommand\@makefntext[1]{\leftskip=0em\hskip0em\@makefnmark#1}
\makeatother
\newcommand\blfootnote[1]{%
  \begingroup
  \renewcommand\thefootnote{}\footnote{#1}%
  \addtocounter{footnote}{-1}%
  \endgroup
}

\providecommand{\keywords}[1]
{
  \small	
  \textbf{\textsf{Keywords---}} #1
}

\renewcommand{\tableautorefname}

\newcommand{\mycomment}[1]{} 
\newcommand*\mytitle{FEQIS: A free--boundary equilibrium solver for integrated modeling of tokamak plasmas}
\newcommand*\myauthor{Fable \textit{et al.}} 
\newcommand*\mycorrmail{emiliano.fable@ipp.mpg.de} 
\newcommand*\myDOI{10.46298/ops.14641} 
\newcommand*\myarxiv{2410.13630} 
 
\newcommand*\myvolume{1}
\newcommand*\myyear{2025} 
\newcommand*\mynumber{2 } 
\newcommand*\mydaterec{October 30, 2024} 
\newcommand*\mydaterevised{December 19, 2024} 
\newcommand*\mydateaccepted{January 27, 2025} 
\newcommand*\mydatepublished{\today}

\newlength{\myshift}
\usepackage[firstpage=true]{background}
\backgroundsetup{contents={\includegraphics[width=1.\paperwidth]{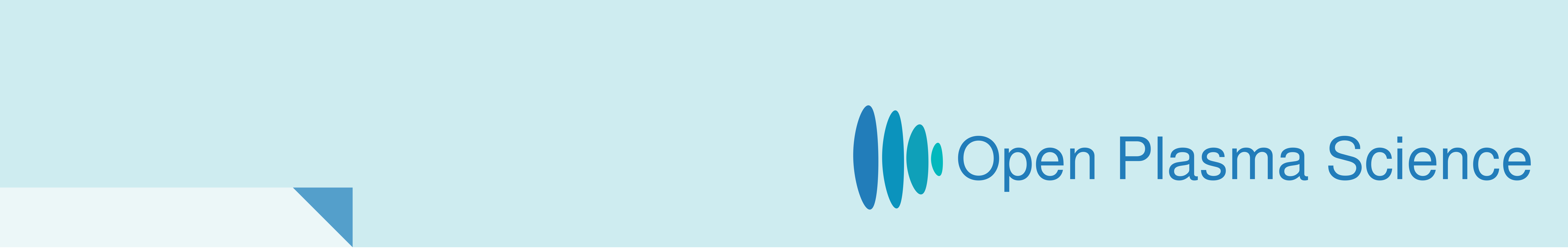}},scale=1,placement=top,opacity=1}

\pretitle{\vspace{-00pt}      \begin{flushleft}   \Large  \color{black} \selectfont  } 
     
\title{\textbf{\mytitle}}

\posttitle{\par \end{flushleft} } 

\preauthor{\vspace{-00pt} \begin{flushleft}  \color{black} \selectfont} 

\author[,1]{Emiliano Fable\orcidlink{0000-0001-5019-9685}$^{*}$}
\author[1]{Giovanni Tardini\orcidlink{0009-0002-0544-6880}}
\author[1]{Louis Giannone\orcidlink{0000-0001-5611-200X}}
\author[a]{the ASDEX Upgrade Team}

\affil[1]{\small Max-Planck-Institut f\"ur Plasmaphysik, Boltzmannstr. 2, 85748 Garching, Germany} 
\affil[a]{see the Author list of H. Zohm \textit{et al}., 2024 \textit{Nucl. Fusion} 64 112001, doi: 10.1088/1741-4326/ad249d}

\postauthor{\par\end{flushleft}} 

\date{}

\begin{document}

\maketitle
\thispagestyle{empty}
\blfootnote{$^*$ Corresponding author: \textsf{\href{\mycorrmail}{\mycorrmail}}}
\blfootnote{Cite as: \myauthor, \mytitle, \textit{Open Plasma Science}  \myvolume , \mynumber (\myyear), doi: \myDOI}


\reversemarginpar

\marginnote{\begin{flushleft}
\sf \bf \footnotesize \color{coolblack} History
\end{flushleft}}[-\myshift]
\addtolength{\myshift}{-0.5cm} 
\marginnote{
\begin{flushleft} \footnotesize
Received \mydaterec
\end{flushleft}}[-\myshift]

\addtolength{\myshift}{-0.5cm} 
\marginnote{
\begin{flushleft} \footnotesize
Revised \mydaterevised
\end{flushleft}}[-\myshift]

\addtolength{\myshift}{-0.5cm} 
\marginnote{
\begin{flushleft} \footnotesize
Accepted \mydateaccepted
\end{flushleft}}[-\myshift]

\addtolength{\myshift}{-0.5cm} 
\marginnote{
\begin{flushleft} \footnotesize
Published \mydatepublished
\end{flushleft}}[-\myshift]

\addtolength{\myshift}{-1cm} 

\marginnote{\begin{flushleft}
\sf \bf \footnotesize \color{coolblack} Identifiers
\end{flushleft}}[-\myshift]
\addtolength{\myshift}{-0.5cm} 
\marginnote{
\begin{flushleft} \footnotesize
DOI \href{https://doi.org/\myDOI}{\myDOI} 
\end{flushleft}}[-\myshift]
\addtolength{\myshift}{-0.5cm} 
\marginnote{
\begin{flushleft} \footnotesize
HAL -
\end{flushleft}}[-\myshift]
\addtolength{\myshift}{-0.5cm} 
\marginnote{
\begin{flushleft} \footnotesize
ArXiv \href{https://arxiv.org/\myarxiv}{\myarxiv} 
\end{flushleft}}[-\myshift]

\addtolength{\myshift}{-1cm} 
\marginnote{\begin{flushleft}
\sf \bf \footnotesize \color{coolblack} Supplementary Material
\end{flushleft}}[-\myshift]
\addtolength{\myshift}{-0.5cm} 
\marginnote{
\begin{flushleft} \footnotesize
- 
\end{flushleft}}[-\myshift]

\addtolength{\myshift}{-1cm} 
\marginnote{\begin{flushleft}
\sf \bf \footnotesize \color{coolblack} Licence
\end{flushleft}}[-\myshift]
\addtolength{\myshift}{-0.5cm} 
\marginnote{
\begin{flushleft} \footnotesize
\href{https://creativecommons.org/licenses/by-nc-nd/4.0/}{CC BY-NC-ND} 
\end{flushleft}}[-\myshift]
\addtolength{\myshift}{-0.5cm} 
\marginnote{
\begin{flushleft} \footnotesize
\copyright  The Authors
\end{flushleft}}[-\myshift]
\addtolength{\myshift}{-0.5cm}

\marginnote{\begin{flushleft} \includegraphics[width=0.55\marginparwidth]{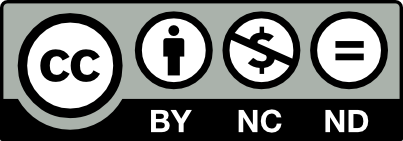}\end{flushleft}}[-\myshift] 

\addtolength{\myshift}{11.75cm} 
\marginnote{\begin{flushleft}
\sf \footnotesize \color{coolblack} \textsc{\textbf{Vol. \myvolume, No. \mynumber (\myyear)}}
\end{flushleft}}[-\myshift]


\addtolength{\myshift}{-1.5cm} 
  
\marginnote{\begin{flushleft}
\sf \normalsize \color{coolblack} \textsc{\textbf{Regular article}}
\end{flushleft}}[-\myshift]

\begin{flushleft}
\textbf{\textsf{Abstract}}
\vspace{5pt} 
\end{flushleft}

A new axisymmetric equilibrium solver has been written, called FEQIS (Flexible EQuIlibrium Solver), which purpose is to be used inside integrated modeling of tokamak plasmas.
The FEQIS code solves the Grad--Shafranov equation and the "circuit" equations for the external coils and passive conducting structures that are toroidally connected.
The code has been specifically equipped with flexibility in choice of circuit connections, and a stripped--down numerical scheme for the solution of the Grad--Shafranov equation through a structure of multi--level simplifications which can be tested against the required accuracy.


\vspace{20pt} 


\begin{flushleft}


\end{flushleft}

\newpage

\newgeometry{ left=20mm, right=20mm,  bottom=3.5cm }

\pagestyle{fancy}
\setlength{\headheight}{15pt}
\fancyhead{} 
\fancyhead[L]{\footnotesize{\href{https://doi.org/\myDOI}{\myDOI}}} 
\fancyhead[R]{\footnotesize{\myauthor}}

\fancyfoot{} 
\fancyfoot[R]{\color{coolblack}{ \thepage \hspace{1pt} | \pageref{LastPage}}}
\fancyfoot[C]{\footnotesize{\color{black} Open Plasma Science \textbf{\myvolume}, No.\mynumber (\myyear)}}

\renewcommand{\headrulewidth}{0.4pt}
\renewcommand{\footrulewidth}{0.4pt}
\renewcommand{\footruleskip}{5pt}
\renewcommand\footrule{\hrule width\textwidth}
\renewcommand\headrule{\hrule width\textwidth}


\tableofcontents


\section{Introduction}

Application of integrated modeling to tokamak full--discharge prediction has in recent times become possible due to the combination of comprehensive transport modeling tools, which describe the dynamics of the thermonuclear plasma, and the integration of said tools into a virtual representation of the actual machine diagnostics, actuation, and control systems. An example of such nested integration is the flight simulator Fenix \cite{fenixjanky1,fenixinterfref,fenixfableref,fenixmodelsmuraca}, which is being developed at ASDEX Upgrade (AUG). In this framework, both the plasma dynamics and the feedback from the control system in response to the diagnosed evolution are modeled in a virtual environment.

From the point of view of simulating the plasma dynamics, one of the basic aspects is the calculation of its magnetostatic equilibrium via the Grad--Shafranov equation (GSE) \cite{gradshafranoveq}, and its evolution on the time scale of induction/resistance with mutual coupling between the plasma and the external conducting coils and structures \cite{blumeqreview}.
The GSE is fundamentally a non--linear equation, requiring a dedicated iteration scheme to converge to a self--consistent solution. 
However, when run inside an integrated modeling framework that is supposed to be fast enough to be run inter--discharge (or even real--time), it would be desirable to find a way to minimize the computational time spent on this problem, especially if massive numerical parallelization is not readily available.

It is noted here that there is an extensive literature on the problem of solving the GSE, both in fixed or in free--boundary mode \cite{mmontecarlogssolverpaper1,fasthighordergssolver2,
	mimeticsegssolverpaper3,accuratederivgssolverpaper4,specelemgssolverpaper5,
	contdynamicsgssolverpaper6}, and sophisticated numerical tools exist that address this problem, such as \cite{TEScodepaper,cedrescodepaper,
	tokamakergscodepaper,createeqcodepaper1}. More recently, there have been very interesting attempts to dramatically speed up the computation by replacing the equation solver with a neural network, e.g. exploiting machine learning and artificial intelligence \cite{nngssolverpinnspaper1,nngssolvergreenpaper2,
	nngssolverfbenetpaper3,nngssolvernaturedegravepaper4}. 
This new line of development has the potential of making GSE solvers extremely fast and still accurate.

In this work we have developed a new equilibrium solver specifically devoted to the problem of being computationally simple and fast with flexibility in retaining or not the more time--consuming aspects such as the non-linear iterations. This new code is presently coupled to the ASTRA transport solver \cite{astracode1,astracode2}. In this work this new code, called FEQIS (Flexible EQuIlibrium Solver), is presented in details, and its application inside the flight simulator Fenix is shown and benchmarked against another established equilibrium solver. 

In Section 2, the new code FEQIS is described in terms of modes of operation and details of some of the novel algorithms. In Section 3, application inside the flight simulator Fenix is shown, comparing it with another equilibrium solver. In Section 4, conclusions are drawn.
\section{Description of FEQIS}
FEQIS comprises 5 main modes of operation, plus the possibility to choose between different options in each mode, which will be described in this and in later sections. 
The 5 main modes of operation are:\\ 1 - Prescribed boundary mode,\\ 2 - Static forward free boundary solution at initialization,\\ 3 - Inverse solution,\\ 4 - Currents dynamics in vacuum,\\ 5 - Full dynamics with plasma.

Moreover, since the code has been specifically written to be used inside fast integrated modeling, focus will be put on the methods to substantially speed it up. At the source, the code is written in Fortran 90 in a modular way, such that it is easy to manage the different algorithms and add new ones. Presently, FEQIS solves the standard GSE, without rotation and without pressure anisotropy. These features will all be added in the future. The possibility of adding ferromagnetic elements is included, however has not been tested yet. The method used for implementing this aspect is that of equivalent magnetization currents. 

Note that the mode 3 (inverse solution) means that a minimization problem is solved to find the set of coil currents that best gives a certain plasma shape. No reconstruction mode is implemented, as done for example in other codes used for real--time equilibrium reconstruction during the experiment itself.
\subsection{Prescribed boundary mode}
First of all, FEQIS can be used as a prescribed boundary Grad--Shafranov solver (PBGSE solver). This means that the plasma boundary, or last--closed--flux--surface (LCFS) is given by the user as a set of $(R,Z)$ points, forming a closed contour, and then FEQIS solves the GSE inside this given boundary, using as inputs the pressure and current profile densities from the core plasma, plus the total plasma current $\ipl$ as a constraint value to rescale the plasma current density $j_\phi$ at each iteration. 

The solver uses an annular grid in non--orthogonal carthesian $(r,\theta)$ coordinates, with $r$ the local minor radius and $\theta$ the carthesian angle. 
The equation $\Delta^* \psi=\mu_0 R j_{\phi}$ is thus solved using a standard LAPACK matrix solver. 
The magnetic axis is adaptively searched for using a least square fit method, such that at convergence, it coincides with the center of the annular grid. Moreover, the radial coordinate grid points at each $\theta$ are moved to coincide with the flux surfaces, thus allowing for easier computation of flux--surface--averaged quantities.

\begin{figure}[!htb]
	\begin{center}
		\includegraphics[width=75mm]{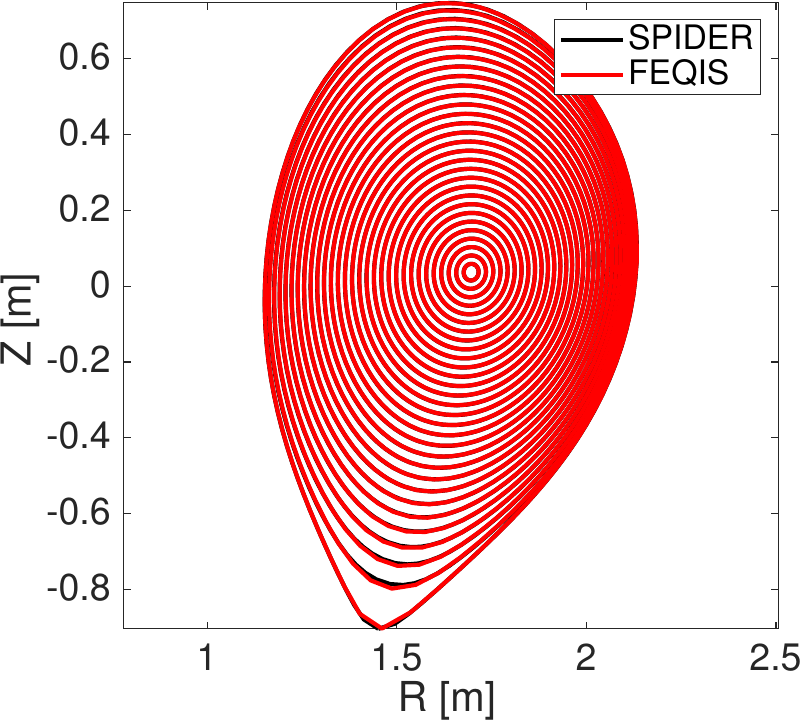}
		\caption{\small Flux surfaces solved by the PBGSE solver, where the diverted plasma boundary is given as input. Black: SPIDER solution, Red: FEQIS solution.}
		\label{figurepbe} 
	\end{center}
\end{figure}

In figure (\ref{figurepbe}) an example of the solution equilibrium for a diverted plasma (AUG $\#34954$ at $t=2.2$ s) is shown.
The comparison shows that FEQIS agrees perfectly with SPIDER \cite{spidercoderef,spidercoderef2} with respect to the distribution of flux surfaces (every closed black/red line represents the same equispaced normalized toroidal flux value).

After convergence, several flux--surface--averaged quantities are computed, that are then sent to an output equilibrium structure and can be used in a 1D transport solver like ASTRA.
\subsection{Free boundary modes}
FEQIS also solves the free--boundary Grad--Shafranov equation (FBGSE), meaning that the magnetic flux boundary conditions are at spatial infinity. For this problem, the Green's function method is promptly employed, to reduce the problem on a closed boundary inside the region of interest \cite{lacknereqcodepaper}. As such, the flux exerted by the coils in the plasma region is computed analytically, and the rectangular grid does not need to encompass all the conducting structures, but can be minimally set around the plasma region (usually just outside the limiter region).

Specifically, the solution grid is Carthesian--rectangular in $(R,Z)$ coordinates, with $R$ being the major radius and $Z$ the vertical axis, in the sense of toroidal cylindrical coordinates, where the toroidal angle $\varphi$ is ignorable, since we focus on axisymmetric systems. 

The FBGSE is solved for using the plasma profiles of pressure and current density, and the external conductor currents. This solution is always "static", as the GSE {\it per se} has no time derivative term, however the equilibration (or convergence) is obtained in different ways depending on the type of problem solved, which is detailed below. 
In addition, also in the free boundary problem, the total toroidal plasma current $\ipl$ is used as a constraint to rescale the plasma current density at every iteration. 
The reason for this choice, and where then the total plasma current is computed, will be explained in more details in subsection (2.3.4). 

In the following subsections, we analyze the individual modes of operation that are under the category of "free boundary problems".
Note that all these modes are not independent in terms of solver usage. In  fact they share the same kernel of solvers, that is the solver for the poloidal flux map is the same, as well as the boundary finding routine and the way the equilibrium profiles are mapped on the 2D flux is the same. The only difference between the various modes shown later is in how the coil currents are treated (static, forward, or inverse computation).
\subsubsection{Static forward solution at initialization}
In this case, the external conductor currents are fixed at their input values (provided by the user), and the solver tries to find a solution using a double iteration strategy.

First, given a guess value of the magnetic axis $(\Rmag^0,\Zmag^0)$, the code finds a vertical and radial stabilization field that sets the plasma in that position, adding on top of the real external field provided by the conductors and the plasma itself.
This step is performed by starting from the generic flux expression:
\begin{eqnarray}
	\psi(R,Z) = \psiplasma(R,Z) + \psiext(R,Z) + \psistabr R^2 + \psistabz Z
	\label{stabfieldscalc1}
\end{eqnarray}
where $\psiplasma$ is the poloidal magnetic flux created by the plasma current density, $\psiext$ the one created by the external conductors, and $R,Z$ the cylindrical coordinates.
$\psistabr,\psistabz$ are the two stabilizing field constants, and the stabilizing field overall is $\psi_{\rm stab} = \psistabr R^2 + \psistabz Z$.

To find the values of the two stabilizing constants, given $AX=(\Rmag^0,\Zmag^0)$, we have to solve for the following pair of  implicit conditions:
\begin{eqnarray}
	\left\vert \frac{\partial\psi}{\partial R}\right\vert_{AX}=0. \nonumber \\
	\left\vert \frac{\partial\psi}{\partial Z}\right\vert_{AX}=0. 
	\label{stabfieldscalc2}
\end{eqnarray}
Let us call $\psiplasma+\psiext = \psi_0$. We can thus expand and invert the system (\ref{stabfieldscalc2}) to obtain the solution formulas:
\begin{eqnarray}
	\psistabr = -\frac{1}{2 \Rmag^0}\left\vert \frac{\partial\psi_0}{\partial R}\right\vert_{AX} \nonumber \\
	\psistabz = -\left\vert \frac{\partial\psi_0}{\partial Z}\right\vert_{AX}
	\label{stabfieldscalc3}
\end{eqnarray}
Obviously, if the plasma have the magnetic axis already coincident with $AX$, the stabilizing fields will turn out to be 0. To compute the gradients in equation (\ref{stabfieldscalc3}) accurately, we expand $\psi_0$ locally around $AX$ employing a 9--points biquadratic interpolant, i.e. $\psi_0 \approx c_1 R^2 Z^2 + c_2 R^2 Z + c_3 R Z^2 + c_4 R Z + c_5 R^2 + c_6 Z^2 + c_7 R +c_8 Z + c_9  $, and use the obtained fit coefficients to compute the gradients analytically.
Note that this process is iterated until the current density map $j_\phi(R,Z)$ is converged (as it depends on the actual flux map including the artificial field).

Secondly, an external Newton iteration scheme tries to send to zero the strength of the stabilizing field $(\psistabr,\psistabz) \rightarrow 0$, by moving around the reference magnetic axis values:
\begin{eqnarray}
	\delta(R,Z)_{AX} = - \mathbf{C}\times (\psistabr,\psistabz) 
	\label{stabfieldsnewtoniter1}
\end{eqnarray}
with $\mathbf{C}$ the 2 X 2 inverse matrix of the Jacobian given by $\partial(\psistabr,\psistabz)/(R,Z)_{AX}$.

When the artificial stabilization field is converged to zero (below a given tolerance), the equilibrium has been found consistently. 
This procedure was already shown to work well in the SPIDER code as described in \cite{spidercoderef,spidercoderef2}. 

\begin{figure}[!htb]
	\begin{center}
		\includegraphics[width=65mm]{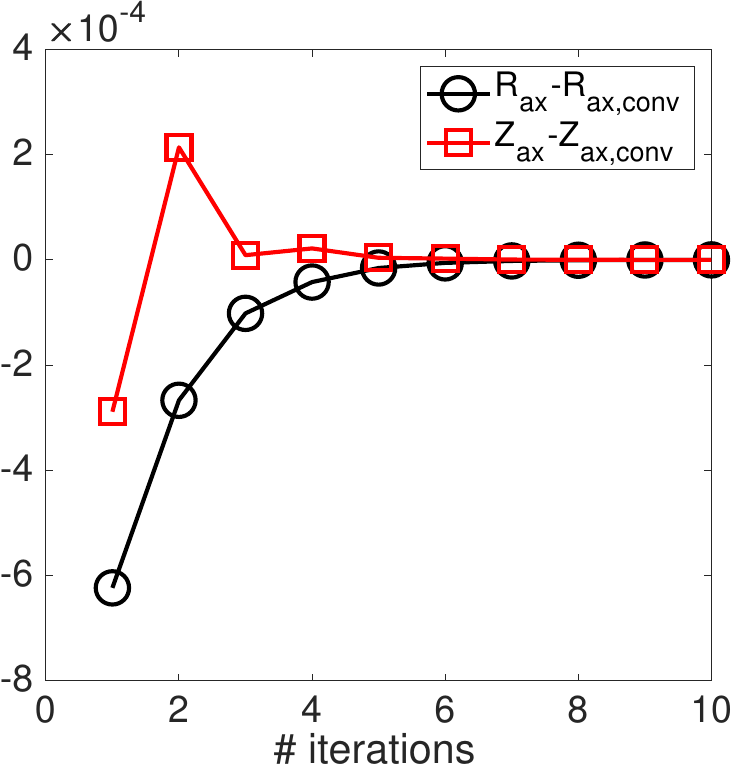}
		\includegraphics[width=65mm]{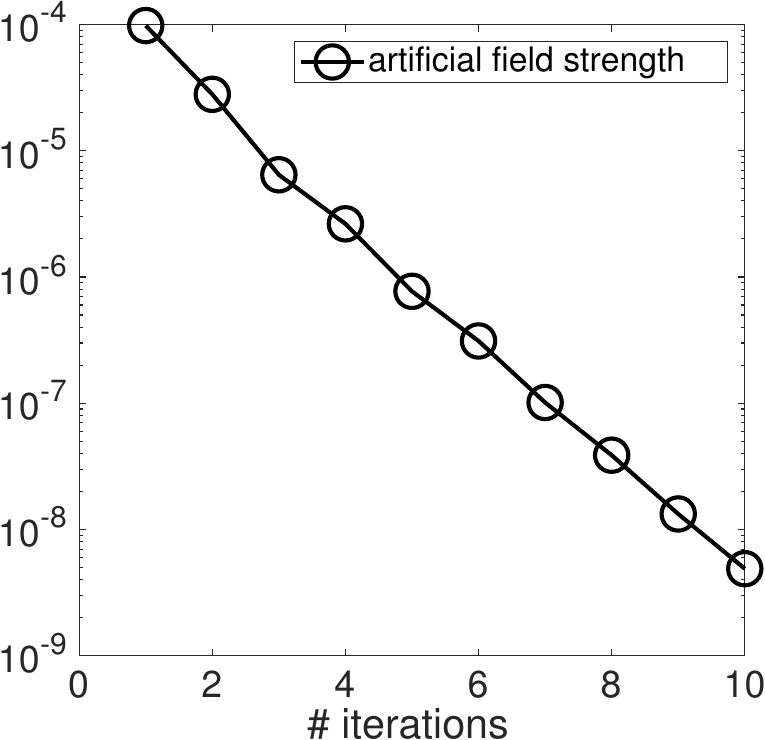}
		\caption{\small Convergence of the first equilibrium using the artificial field technique: (left) magnetic axis major radius $R_{\rm ax}$ and vertical position $Z_{\rm ax}$ relative to the converged value as a function of the iteration nr.; (right) strength of the artificial field $\xi_{\rm stab}$.}
		\label{figurefbestaticinit} 
	\end{center}
\end{figure}

In figure (\ref{figurefbestaticinit})(left) the trajectory of the values of $R_{\rm ax},Z_{\rm ax}$ is shown along the number of iterations at the first equilibrium call, with the sum of the artificial field squared $\xi_{\rm stab}=\sqrt{\psistabr^2+\psistabz^2}$ on the (right) plot.
For this specific case, the same plasma shown in figure (\ref{figurepbe}), but run in free boundary mode, the first equilibrium converges after 10 iterations, where the tolerance has been set as $\xi_{\rm stab} < 10^{-8}$.
\subsubsection{Static inverse solution at initialization}
If the coil currents do not have to be strictly fixed, but can be varied a little to allow the plasma to find an equilibrium arbitrarily close to the one given as guess (for example prescribing a boundary and a magnetic axis), then this inverse mode performs this task. In this case, a global optimization problem is solved, where the cost function is given as:
\begin{eqnarray}
	 F_{\rm gloptim}=\sigma_B \sum_b\left(\psi_b - \langle\psi_b\rangle \right)^2+ \sum_j \sigma_{I,j} \left(I_j-I_j^0\right)^2+\sigma_{ax}\left(\left\vert\frac{\partial \psi}{\partial R}\right\vert_{ax}^2+\left\vert\frac{\partial \psi}{\partial Z}\right\vert_{ax}^2\right)
	\label{costfunctiongsoptim}
\end{eqnarray}
where $\sigma_B,\sigma_{I,j},\sigma_{ax}$ are weight coefficients, respectively for the reference boundary points, the reference conductor currents, and the reference magnetic axis position (which does not need to coincide with the real magnetic axis). $\psi_b$ is the actual value of the magnetic flux on the prescribed boundary points (index "b", running from 1 to $N_b$), and $\langle\psi_b \rangle$ is their average. $I_j,I_j^0$ are respectively the conductor currents and their reference values of the conductor $j=1..N_{\rm conduc}$, and finally the last term is the amplitude of the magnetic flux gradient on the reference axis position (while it being strictly 0 on the true magnetic axis). 
Note that the various weight coefficients $\sigma$'s do not need to have the same units (and as such similar values). However it is not difficult to identify possible normalization as a future task.

To simplify the problem characterized by the cost function (\ref{costfunctiongsoptim}) and make it linear, one first expresses the full flux as $\psi = \psi_{\rm plasma} + \psi_{\rm ext}$, with the former the flux produced by the plasma current density and the latter the flux produced by the external conductors. We then make the approximation that $\vert\partial\psi_{\rm plasma}/\partial I \vert\ll\vert\partial\psi_{\rm ext}/\partial I \vert$, where $I$ is a generic conductor current. If the plasma was a non--deformable conductor, the left--hand--side of the comparison would be strictly 0. However, even with the plasma as a fluid, that approximation is still valid especially close to the actual solution. 

To solve the optimization problem, one first computes analytically the derivative of the cost function with respect to the conductor current $j$, which is:
\begin{eqnarray}
	\delta F_{\rm gloptim,j}=2 H_j \nonumber \\
	H_j = \sigma_{I,j} \left(I_j-I_j^0\right) +\sigma_B \sum_b\left[\left(\psi_b - \langle\psi_b\rangle \right)(G_{j,b}-\langle G_{j,b}\rangle)\right]+ \nonumber \\
	+\sigma_{ax}\left(\left\vert\frac{\partial \psi}{\partial R}\frac{\partial G_j}{\partial R}\right\vert_{ax}+\left\vert\frac{\partial \psi}{\partial Z}\frac{\partial G_j}{\partial Z}\right\vert_{ax}\right)
	\label{costfunctiongsoptimderiv}
\end{eqnarray}
where $G_j$ is the Green function from the conductor $j$ to the boundary point $b$ or to the magnetix axis "ax".

The Newton iteration scheme in this case is the following:
\begin{eqnarray}
	\delta I_j = -(\mathbf{M} \times \delta F_{\rm gloptim})_j
	\label{costfunctiongconvnewtscheme}
\end{eqnarray}
where $\mathbf{M}$ is a matrix obtained in this way:
\begin{eqnarray}
	\mathbf{M}=(2X)^{-1} \nonumber \\
	X_{i,j}=\sigma_{I,i}\delta_{i,j}+\sigma_B\sum_b\left(G_{i,b}-\langle G_{i,b}\rangle \right)\left(G_{j,b}-\langle G_{j,b}\rangle \right) 
	+\sigma_{\rm ax}\left(\frac{\partial G_i}{\partial R}\frac{\partial G_j}{\partial R}+\frac{\partial G_i}{\partial Z}\frac{\partial G_j}{\partial Z}\right)_{\rm ax} 
	\label{costfunctiongconvnewtschemematrix}
\end{eqnarray}
where $\delta_{i,j}$ is the Kronecker symbol $\delta_{i,j}=1$ if $i=j$, 0 otherwise.
Since the matrix $X$  is independent of the magnetic flux or the conductor currents, it can be pre--computed before the iterations are performed. 

When the iteration scheme, equation (\ref{costfunctiongconvnewtscheme}), converges, the correction terms $\delta I_j$ are applied to the conductor currents to find the new vacuum magnetic flux on which the plasma develops.

It is also possible to compute the coil currents from scratch, that is with null reference values. In this case, the cost function (\ref{costfunctiongsoptim}) is modified in that the coils term is:
\begin{eqnarray}
	F_{\rm gloptim}=...+ \sum_j \sigma_{I,j} \frac{1}{2} L_j I_j^2+...
	\label{costfunctiongsoptim2}
\end{eqnarray}
where $L_j$ is the self--inductance of conductor $j$. This makes the cost function sensible on the magnetic energy generated by each individual coil.

Finally, inspired by what has been implemented in the NICE code \cite{nicepaperref}, the code can also compute a set of currents, which develop a time trajectory of provided shapes, including the required loop voltage that the plasma needs to maintain its plasma current $\ipl$. For this case, additionally the plasma external inductance $\Lext$, plasma current at the present time slice $\ipl$, and the differential variation of the plasma boundary flux during the lapsed time $\delta\psi = \Vloop\delta t + \Lext \delta\ipl$. The cost function is then modified by the following additional term:
\begin{eqnarray}
	F_{\rm gloptim}=...+\lambda\left[\delta\psi-\Vloop\delta t-\Lext \delta \ipl \right]
	\label{costfunctiongsoptim3}
\end{eqnarray}
where $\lambda$ is a Lagrange multiplier. This additional term enforces the flux balance between the two time slices, where the external flux differential created by the coils has to balance the flux loss by plasma resistance and inductance.
Note that voltage limits for each coil can be given, in which case the optimal value of the new coil currents have to fall in between the two current limits given by the estimate:
\begin{eqnarray}
	I_{\rm lim}=I_{\rm ref} + \delta t \left[V_{\rm lim}-R I_{\rm ref}  \right]/L
	\label{costfunctiongsoptim31}
\end{eqnarray}
where "ref" means the coil current found at the previous time point, and $R,L$ are respectively the resistance and self--inductance of the coil. These constraints maintains the time variation of the fitted coils smooth and sensible even if the actual circuit equations are not really solved.

\begin{figure}
	\begin{center}
		\includegraphics[width=65mm]{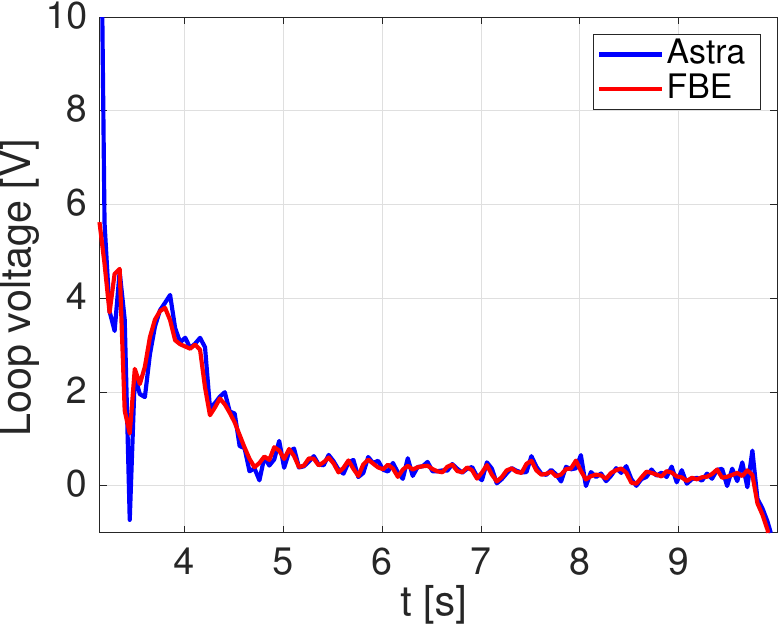}
		\includegraphics[width=85mm]{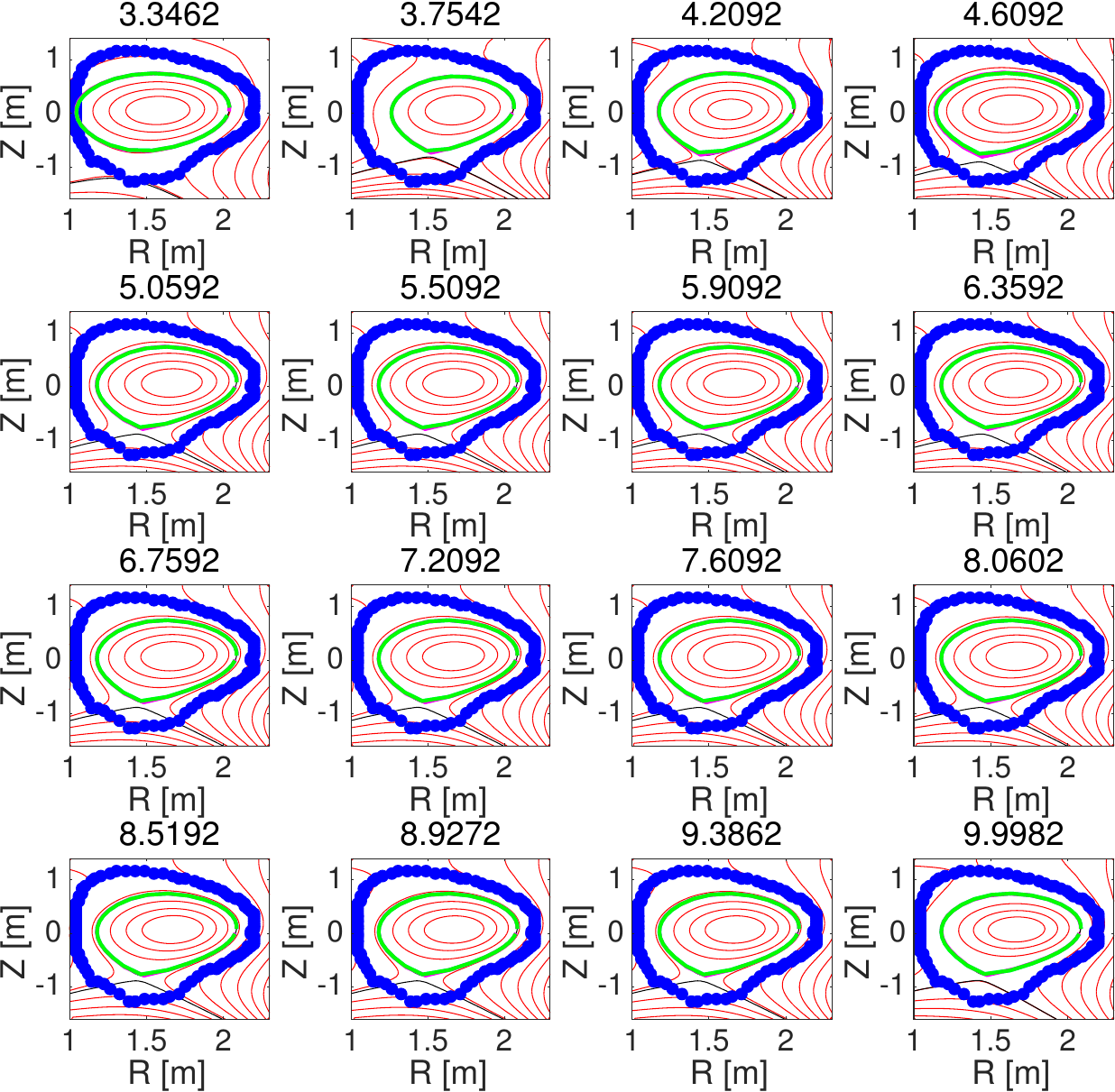}
		\caption{\small (left) Comparison of the loop voltage estimated in Astra to get the requested plasma current given the plasma resistivity (blue line, "Astra") and the loop voltage calculated a--posteriori from the fitted coil current evolutions, shown in figure (\ref{figurefbeinversecoilfitdynamic2}) (red line); (right) comparison of the requested shapes (green) and the fitted shapes (black line, whereas the red lines are all the other $\psi$ contour lines). The blue thick contour represents the limiter surface used in the simulation. When the plasma touches this contour, it becomes limited, otherwise it is an X-point configuration.}
		\label{figurefbeinversecoilfitdynamic} 
	\end{center}
\end{figure}

In figure (\ref{figurefbeinversecoilfitdynamic}) an example of dynamical coil fitting for an entire AUG discharge ($\#40446$) is shown. 
\begin{figure}
	\begin{center}
		\includegraphics[width=125mm]{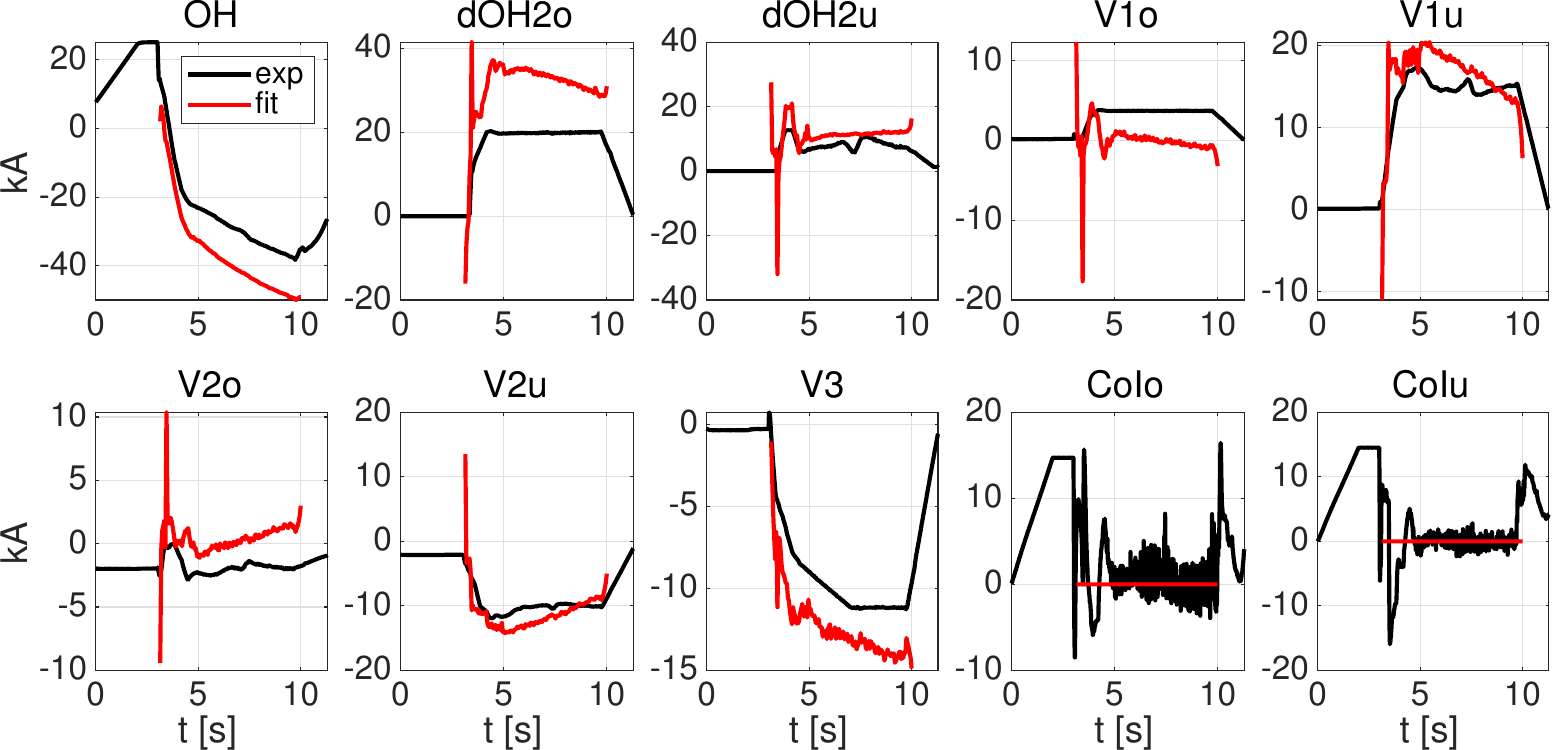}
		\caption{\small Comparison of the fitted coil currents "fit" (red lines) with the experimental currents "exp" (black), for the active coils of AUG: OH, dOH2o, dOH2u, V1o, V1u, V2o, V2u, V3. The position control coils CoIo and CoIu have not been fitted.}
		\label{figurefbeinversecoilfitdynamic2} 
	\end{center}
\end{figure}
The left plot shows the comparison between the loop voltage calculated from Astra (blue line) to get the desired plasma current (imposed from the experimental time trace) and the one calculated a--posteriori from the fitted coil currents evolution (red curve). The comparison shows that, as desired, the fit procedure satisfies the constraint imposed on the change of external magnetic flux, such as to balance the required loop voltage at the plasma boundary. 

On the right plot, the comparison of the fitted separatrix (black) with the requested shape (green) is displayed. The fit maintains the requested shape almost to perfection. 

Finally, in figure (\ref{figurefbeinversecoilfitdynamic2}), the fitted coil currents (red) are compared with the experimental coil currents (black). Note that the fit procedure knows nothing about the experimental currents. For this case, the fit constant $\sigma$s have been chosen to come close to the measured coil currents, but a discrepancy is expected since the fit does not know about current limits nor voltage limits (which have not been used in this case). Note also that the CoIo and CoIu currents, which are active control currents, have not been used for the fit, since these coils are only used for active control of the plasma during operation, and are not envisioned as static shaping coils for designing shapes in AUG. 
\subsubsection{Dynamical evolution in vacuum}
For the external conductors, both active and passive, FEQIS solves circuit equations of the type:
\begin{eqnarray}
	L_j \frac{d I_j}{dt} + \sum_i \left[ M_{i,j}\frac{d I_i}{dt}+R_{i,j}I_i\right] = V_j + {\rm (plasma \; term)}_j   
	\label{circuiteqtype1}
\end{eqnarray}
with $L_j,M_{i,j}$ the self and mutual inductances, and $R_{i,j}$ the resistance matrix (allowing for complex connections of the coils). The supplied voltage to the coils is $V_j$, while passive conducting structures are assigned 0 external voltage input. The last term on the right--hand--side is the term that represents the effect of the magnetic field produced by the plasma onto the conducting structures.

Note that the code is also capable of changing the circuit connections (resistance matrix, inductances, and number of coils) during the time evolution, if required, which reflects the case of some devices where resistors can be switched on or off, and some coils can be switched on or off during operation. It is also to be stressed that only axisymmetric circuits and structures are represented in this code. 3D passive elements or closed loop currents that do not close toroidally cannot be taken into account except via some "effective" resistance effectively acting in the toroidal direction.

This system of 0D (in space) equations is solved in time with an implicit scheme, to ensure stability even at large time step. The only explicit term being the plasma term, thus requiring the time step being at least smaller than the typical time scale of plasma motion due to the varying conductor currents.  In the case of vacuum calculations in absence of the current--carrying plasma, the "plasma term" is set to 0.
\subsubsection{Dynamical evolution with plasma}
After the gas is ionized at the breakdown inside the vacuum chamber, the equations (\ref{circuiteqtype1}) start to include a finite plasma term:
\begin{eqnarray}
	{\rm (plasma \; term)}_j = -\frac{d\Psi_j}{dt}  
	\label{circuiteqplasmaterm}
\end{eqnarray}
with $\Psi_j$ the magnetic flux created by the plasma onto the region of conductor $j$.

Since the plasma flux is computed by the static GSE, the system comprising of equations (\ref{circuiteqtype1}) and GSE are iterated at fixed time slice until the sum of conductors and plasma fluxes converges. Note that this iteration scheme converges only if the plasma+coils system is in the "resistive" branch of the MHD spectrum. Alfv\`enic dynamics is not included in this framework.
\subsection{Details on the algorithm to solve the FBGSE}
Independently of the mode of operation chosen to solve the free boundary equation, a few steps are always performed to calculate the full solution. Here the details of these steps are presented.
\subsubsection{Rectangular GSE solver}
Given the toroidal current density distribution $j_\phi(R,Z)$ and the external flux map $\psiext(R,Z)$, the full flux is obtained by linearly overlapping the plasma flux $\psi_{\rm pl}$ and $\psiext$. The plasma contribution is obtained by solving the following 2 steps system inside the simulation domain $(R,Z)\in\Omega$ defined as $\Omega$, whose boundary is called $\partial\Omega$:
\begin{eqnarray}
	\Delta^* g = j_\phi      \;\;\;\;\;\; ; \;\;\;\;\;\; \left[g\right]_{\partial\Omega}=0   \nonumber\\
	\Delta^* \psi = j_\phi      \;\;\;\;\;\; ; \;\;\;\;\;\; \left[\psi\right]_{\partial\Omega}=z(g)  
	\label{plasmagsesolver}
\end{eqnarray} 
where the boundary function $z(g)$ is obtained by performing boundary integrals of the Green function and the perpendicular gradient of $g$. Note that the cyclic integral, when it happens to consider self--inductance of a small segment of boundary, uses an analytical formula for the segment self--inductance as computed in \cite{lacknerseginductanceformula}.
Numerically, the pair of equations (\ref{plasmagsesolver}) are solved using the discrete sine transform method in the vertical direction as in \cite{sintransformeqgarching}.

Once the plasma solution is obtained, the full flux is given by: $\psi=\psi_{\rm pl}+\psiext + (artificial \; field \;  if \; needed)$.
\subsubsection{Finding the value of the magnetic axis position and the LCFS flux}
Now one has the full map $\psi(R,Z)$. To find the magnetic axis, one starts from the previously found value (or initially from a guess value), and a simple "uphill" search algorithm will find the grid point with the maximum value of $\psi$. After this, a local 9--point exact biquadratic interpolation routine will refine the solution and find the real values of $\Rmag$ and $\Zmag$. In practice, upon expanding locally $\psi \approx c_1 R^2 Z^2 + c_2 R^2 Z + c_3 R Z^2 + c_4 R Z + c_5 R^2 + c_6 Z^2 + c_7 R +c_8 Z + c_9$, a Newton scheme is employed to find the location that satisfies $\partial\psi/\partial(R,Z)=0$. 

For the plasma boundary value, $\psi_b$, the algorithm is a bit more complex. At the very first calculation, the code performs a full sweep of the 2D grid, and finds all points that satisfy the condition $\displaystyle \left\vert\frac{\partial \psi}{\partial R}\right\vert^2+\left\vert\frac{\partial \psi}{\partial Z}\right\vert^2=0$, which can be either O--points or X--points (aka null--gradient--points or NGPs). Obviously, the O--point which coincides with the magnetic axis is discarded as it cannot be the plasma boundary at the same time. Next, all NGPs that lie outside the limiter region are excluded.
Conversely, all limiter points that are in the shadow of an X--point are discarded. This is done by considering the domain divided in 4 quadrants with respect to the magnetic axis. Any X--point point that appears in each quadrant defines the maximum (or minimum) side of the box in which the plasma should be contained. In this way, it is clear that the plasma LCFS cannot bypass an existing X--point on the same side (right, left, top, bottom). This is a rather crude description of the X--point shadow regions, but it is found to be working well for typical tokamak plasma configurations. 
Finally, all NGPs that are non--monotonically connected to the magnetic axis are also excluded. That is, if going from the NGP to the magnetic axis, there is an inversion of the magnetic flux gradient, the NGP cannot be the plasma boundary. Moreover, limiter points that are in the shadow of NGPs are also not considered. This is simply achieved by checking in which quadrant, with respect of the magnetic axis, the NGP point is, and then cutting the domain on the other side of the NGP point, with respect to the magnetic axis. This is a rather simple and crude procedure, but it works for typical tokamak equilibria.

After having removed all pathological flux grid points, what remains between limiter points and NGPs is compared in terms of the magnetic flux: the highest magnetic flux is assigned as the real plasma boundary value.
From the second iteration on, or when the dynamical calculations are performed, the full 2D sweep is not done. Instead, a sweep is done over the plasma LCFS of the previous iteration, to find (eventually) new NGPs.
Note that the search algorithm uses again a 9--points exact biquadratic interpolant, where the null--gradient condition is obtained via a local Newton method solution.

When the poloidal flux value for the plasma boundary is found, an algorithm is employed to define a set of points that describe this boundary in polar coordinates $(\psi,\theta)$ (Carthesian $\theta$) which is then used in prescribed boundary mode to evaluate the internal flux surface with more accuracy and compute the flux--surface--average geometric quantities eventually needed by the transport code that embeds FEQIS. This algorithm moves along the angle $\theta$, and for each angle, moves along the ray from the magnetic axis outwards, in steps of $ds=\sqrt{(dr \cos\theta)^2+(dz \sin\theta)^2}$, with $dr,dz$ the grid resolution steps in radial and vertical direction.
\subsubsection{Deploying the new map of the current density}
Once the new magnetic axis and LCFS flux values are found, the flux $\psi$ is normalized between 0 (axis) and 1 (boundary). 
Next, a sweep algorithm, which starts from the closest grid point to the magnetic axis, checks all grid points moving outside in spiraling fashion, until either the boundary flux is found, or the NGP or limiter point is found.
Points close to the boundary, but in the exterior vacuum region, are also stored as "ghost" points, which will be discussed briefly later on.
The plasma current density is assigned to the interior points simply by interpolating the values of $P',FF'$ from their original $\psi$ normalized grid, onto the free--boundary solution normalized $\psi$.

Since the boundary can cut in between grid points in different ways, an algorithm is used to assign a current density to the exterior "ghost" points, to mimic the fact that the plasma partially occupies cells. 
Suppose that the boundary would cut between grid points $1G$ and $2G$, where "1" is at the left of "G", and "2" is under "G". Let us call $t_1$ and $t_2$ the normalized distances between $1B_{1G}$ and $2B_{2G}$, where $B_{1G},B_{2G}$ are respectively the location of the intersection between the real boundary and the grid points connection grid lines.
Let us also call the values of the current densities at $B_{1G},B_{2G}$ as $j_{1G}$ and $j_{2G}$. Then the value of the assigned current density on the ghost external point $G$ is: $\displaystyle j_G = t_1 j_{1G} + t_2 j_{2G}-t_1 t_2 (j_{1G}+j_{2G})/2.$.
Then, we also define $\displaystyle I_G = t_1+t_2-t_1 t_2$ , whereas $I_G=1$ for interior points. The current is then smoothed by employing a second order Shapiro filter for each grid point $(l,k)$: $\displaystyle j_{\rm corrected}(l,k)=I_G(l,k) 0.5 \left[j(l,k)+0.25(j(l+1,k)+j(l-1,k)+j(l,k+1)+j(l,k-1)) \right]$. This smoothing avoids extreme current profile gradients caused by a coarse--gridding of the 1D profiles.
It can be seen that the formula for the ghost points satisfies all the properties of the current density whenever the boundary intersects an actual grid point or when either $t_1,t_2$ are 0.

This way of distributing the edge current density on external ghost points makes the code physically sensitive to the boundary moving in between grid points, which is one goal of this method. The accuracy of this method is however low order (first order accuracy), whereas other methods can be found in the literature that reach second order accuracy \cite{fbeextrapcurrentspasche}. These more sophisticated methods could be implemented and tested in the future.
\subsubsection{Coupling with the 1D current diffusion equation}
The 2D GSE solution provides only the geometry of the flux surfaces, both for the nested closed field line region (plasma core) and the external region which is characterized by open field lines and the presence of the external conductors.
In the plasma core, from the side of the transport solver, a 1D current diffusion equation (CDE) is usually solved for, to provide the 1D profile of the poloidal magnetic flux $\psi$ as a function of the underlying radial coordinate, which usually is the toroidal magnetic flux $\Phi$. As such, the 1D CDE provides $\psi_{1D}(\Phi)$, whereas the 2D FBGSE provides $\psi_{2D}(R,Z)$. From the point of view of the code evolution, the two fluxes are independent of each other. However, at the interface (the LCFS), the two fluxes must have the same value: 
\begin{eqnarray}
	\psi_{1D}(LCFS)=\psi_{2D}(LCFS)
	\label{psilcfscond1}
\end{eqnarray}
This condition has to be forced from the side of the plasma core. That is, equation (\ref{psilcfscond1}) has to be seen as a definition of $\psi_{1D}(LCFS)$, that is the boundary condition for the CDE. 

The problem is then to find an expression of $\psi_{2D}(LCFS)$, that contains implicitly $\psi_{1D}(LCFS)$, since equation (\ref{psilcfscond1}), when taken as an explicit definition, is extremely unstable during the time evolution.
To solve this issue we follow \cite{dinarefpaper} and employ the following exact relation:
\begin{eqnarray}
	\psi_{1D}(LCFS)=\langle \psiext \rangle_b + \langle \psi_{\rm plasma} \rangle_b
	\label{psilcfscond2}
\end{eqnarray}
where $\langle ...\rangle_b$ denotes averaging over the plasma boundary (LCFS), and the
second term on the right hand side represents the plasma contribution to the boundary flux. This contribution can be represented as $\langle \psi_{\rm plasma} \rangle_b = L_{\rm ext}\ipl$, with $L_{\rm ext}$ the external inductance of the plasma, and $\ipl$ the plasma current. The value of the external inductance is calculated via a double integral over the boundary using the Green function: $\displaystyle L_{\rm ext}=\frac{1}{2\pi\mu_0\ipl}\langle \oint \frac{1}{R}G \vert \nabla{\psi}\vert dl \rangle_b$. On the other hand, the plasma current is expressed through the boundary gradient of the poloidal flux: $\ipl = G (\partial\psi/\partial\rho)_b$, with the geometrical parameter $G$ given by:
\begin{eqnarray}
	G =\frac{1}{4 \pi^2 \mu_0}\frac{dV}{d\rho}\left\langle\frac{\vert\nabla\rho\vert^2}{R^2}\right\rangle
	\label{lextformula1}
\end{eqnarray}
with $V$ the local plasma surface volume and $\rho$ a generic radial coordinate (in ASTRA it is defined as 
$\rho=\sqrt{\Phi/(\pi B_0)}$ with $\Phi$ the toroidal magnetic flux and $B_0$ the reference toroidal magnetic field). In the following, we omit the "1D" pedix of the flux $\psi$ calculated in the current diffusion equation except when said otherwise.

Physically, condition (\ref{psilcfscond2}) simply says that the total flux on the plasma boundary is the sum of the vacuum flux created by external conductors plus the flux generated by the plasma itself. Substituting equation (\ref{psilcfscond2}) in (\ref{psilcfscond1}) we find our implicit boundary condition for the CDE:
\begin{eqnarray}
	\psi_b=\langle \psiext \rangle_b + L_{ext} G \left\vert\frac{\partial \psi}{\partial \rho}\right\vert_b
	\label{psilcfscond3}
\end{eqnarray}
where the external flux produced by the coils and the conducting structures $\psiext$ acts effectively as a source of boundary flux.
This implicit (and thus fully stable) boundary condition determines simultaneously the values of $\psi_b$ and $\ipl$ and lead to the full closure of the dynamical problem.
\subsubsection{Speeding up the code}
As mentioned in the introduction as a well known fact, the FBGSE system is non--linear, because the plasma current density is remapped from the flux surfaces $\psi$ of the old iteration, to the new iteration result. As such, one single calculation does not suffice. Moreover, the kinetic profiles themselves need to be iteratively re--adapted on the new flux surfaces. This requires performing nested iteration cycles if the numerical scheme does not solve the entire plasma system in one single place (e.g. transport solver + equilibrium solver usually solve on different grids).
Performing the full iteration scheme may be rather time--consuming, especially if parallelization of the individual elements is not available. 
For this reason, in FEQIS the user has the option to switch off any sort of iteration scheme at fixed time steps, and instead use the time variable as effective "iteration parameter". This is justified if the time step is lower than the typical time scales like the confinement time, or the current density/plasma motion (resistive time), or the conductor currents (L/R) time scale. 

Moreover, the calculation of the conductor currents is decoupled from the plasma term itself. That is, in equation (\ref{circuiteqtype1}), the plasma term, equation (\ref{circuiteqplasmaterm}), is "frozen" in between two consecutive GSE calls, where the time derivative is correctly computed using the equilibrium calls time difference.
In this way, while the vacuum field is effectively changing following the conductor currents evolution, the plasma itself is only updated every now and then. This is particularly advantageous in steady--state phases with no particularly strong perturbations. The algorithm is devised to call the GSE when relative changes in several parameters (defined by the user) are observed, e.g. plasma $\beta$ and $l_i$, or the plasma position $(R,Z)_{AX}$.
The non--linear iterations can be switched on at any time during the code run, thus allowing the user to accelerate or slow down the calculation depending on the accuracy needed. For example, the non-linear iterations could be switched back on if a vertical displacement event (VDE) is detected (check on the vertical position variation). Presently, no specific criterion is included in the code, whereas this has to be defined by the user itself. In the simulations shown later, the criterion used to define at which time the GSE solver is called (non--linear iterations are permanently switched off) is described below. First, the time between calls can only vary between a minimum which is the same as the circuit equations $\tau_{\rm cirq}$, and a maximum which we define as $5\tau_{\rm cirq}$. Starting from the minimum, the time between calls is increased by $10\%$ at every time step if $\varepsilon<1\%$, where $\varepsilon=(\vert Z-Z_0\vert+\vert R-R_0 \vert)/a$, and the magnetic axis position values $R,Z$ are the new evaluation and $R_0,Z_0$ is the old evaluation. $a$ is the plasma minor radius. If the error $\varepsilon>1\%$, the interval between calls is reduced by $20\%$.

Finally, the code can be run with the minimal grid resolution compatible with still accurate results, which in the simulations shown later, is 65x65 grid points, and the solution region is minimally surrounding the limiter area.
A comparison of the results obtained using a 65x65 or a 129x129 grid will be shown later in figure (\ref{figure5}).
\section{Application in Fenix and benchmark against SPIDER}
The FEQIS code has been coupled to the ASTRA transport solver \cite{astracode1,astracode2}, and used as the dynamical FBGSE solver to carry out full--discharge simulations in the flight simulator Fenix \cite{fenixfableref}. Here we apply this integrated modeling tool to ASDEX Upgrade discharge $\#40405$, a H--mode discharge characterized by a short flat-top and long ramp--down.
ASTRA is also coupled to the free boundary GSE solver SPIDER \cite{spidercoderef2}, which is used here to benchmark the new code FEQIS.

\begin{figure}
	\begin{center}
		\includegraphics[width=150mm]{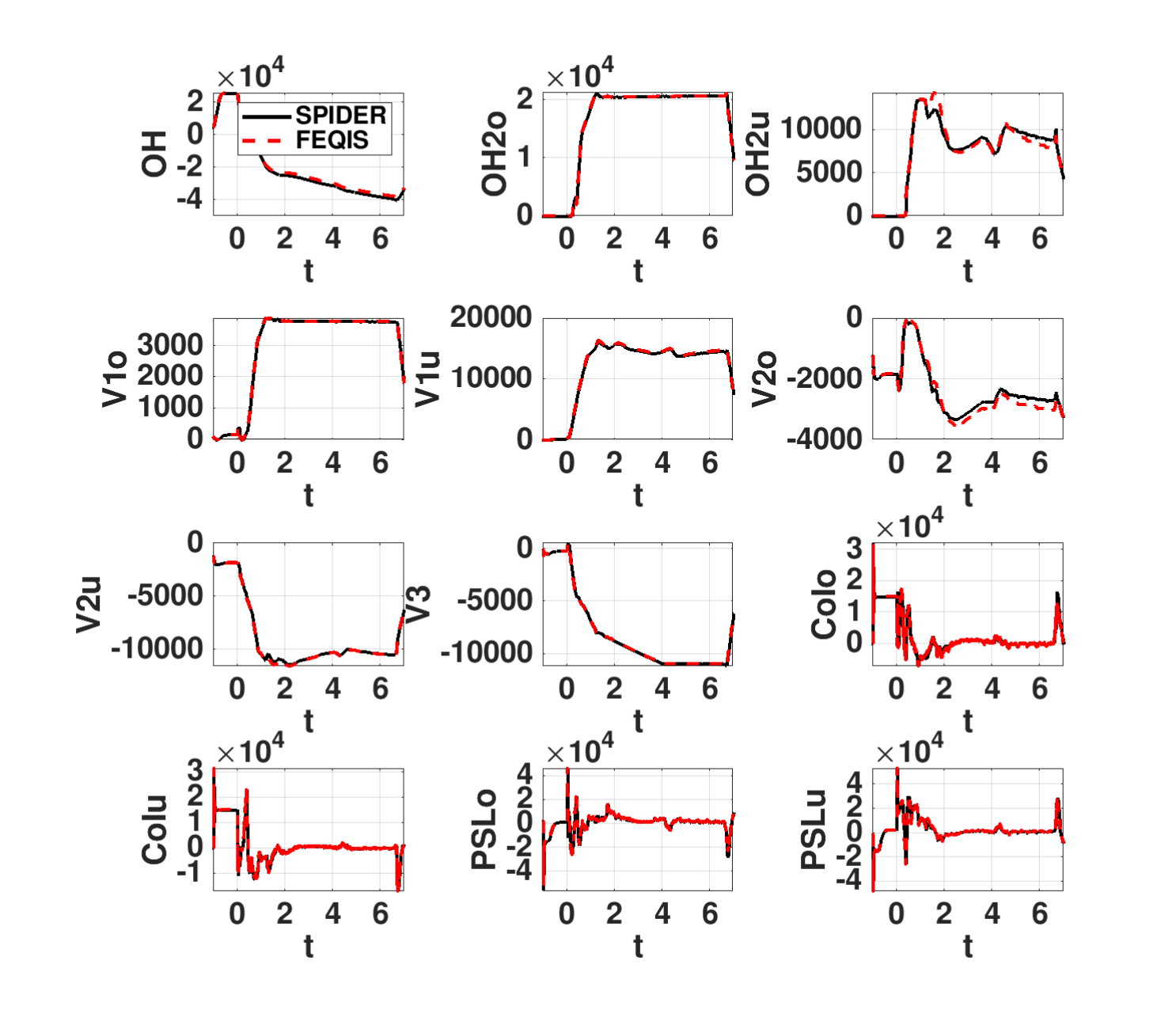}
		\caption{\small Time traces of the various conductor currents, all given in units of A/turn, as a function of the time along the discharge, where $t=0$ marks the start of the breakdown plasma initiation and $t<0$ is the vacuum phase. The code results using SPIDER are in black, whereas the code results using FEQIS are in red.}
		\label{figure1} 
	\end{center}
\end{figure}

\begin{figure}
	\begin{center}
		\includegraphics[width=150mm]{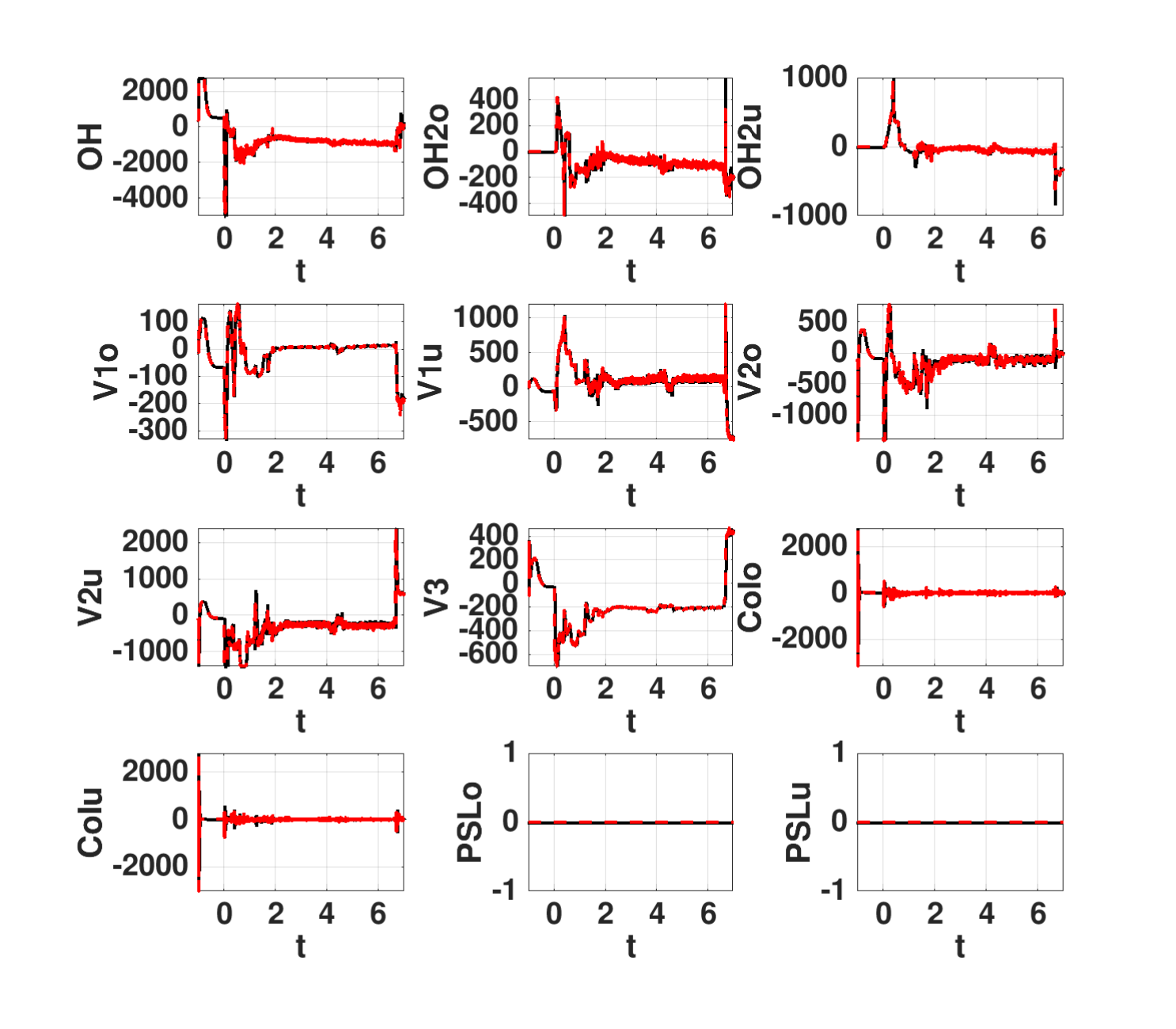}
		\caption{\small Time traces of the voltage applied to each individual conductor. The voltage is given in units of V. For the last two coils PSLo and PSLu, which are passive stabilizing coils, no external voltage is supplied. Color coding is the same as in the previous figure.}
		\label{figure2} 
	\end{center}
\end{figure}

\begin{figure}
	\begin{center}
		\includegraphics[width=170mm]{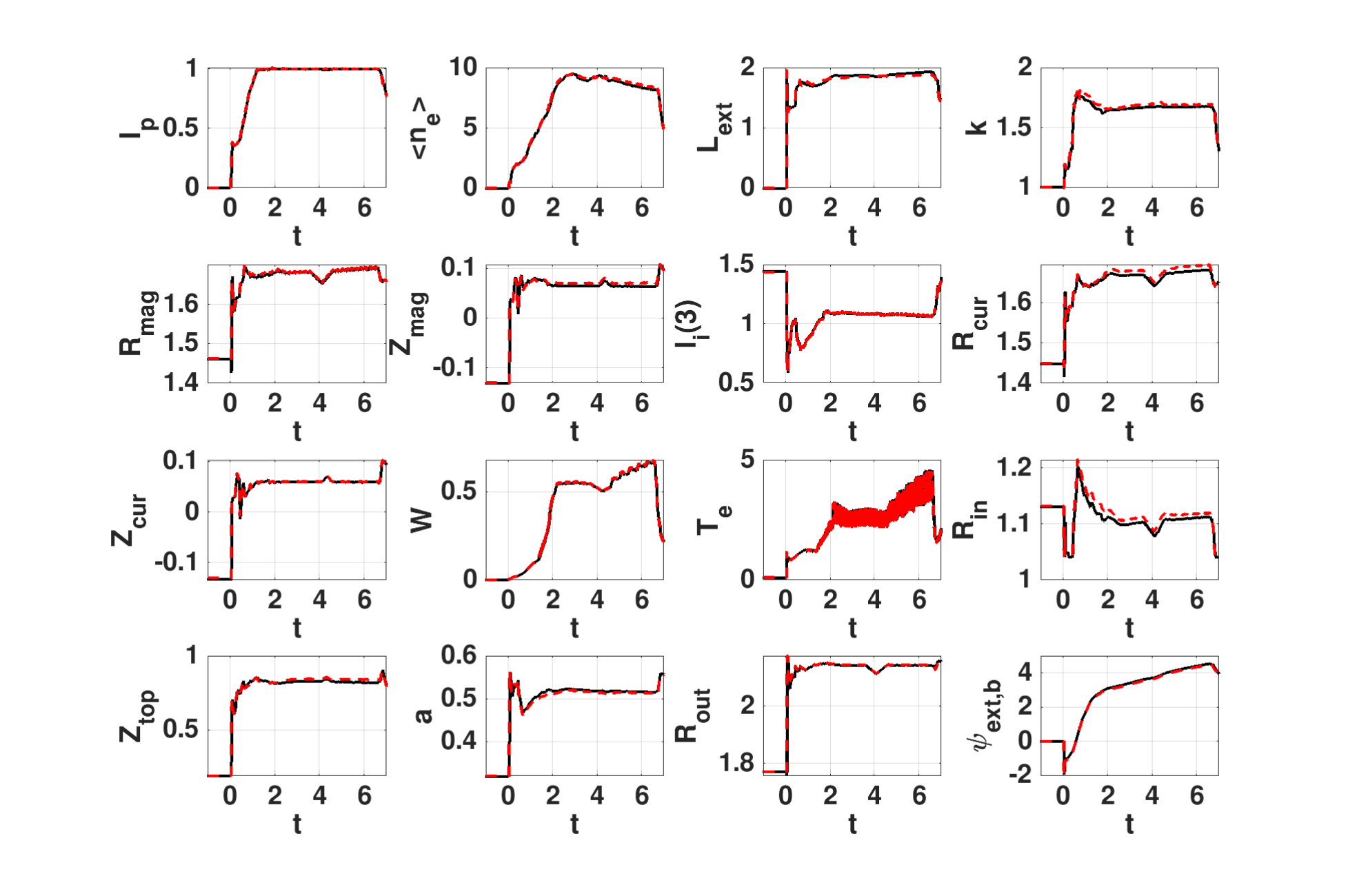}
		\caption{\small Time traces of various quantities: plasma current $\ipl$ [MA], line averaged density $\langle \dense \rangle$ in $10^{19} m^{-3}$, plasma external inductance $L_{\rm ext}$ normalized to $2\mu_0 R$, plasma elongation $k$, outer strike point vertical position (o.s.p.) in [m], inner strike point vertical position (i.s.p.) in [m], internal inductance $l_i(3)$, current centroid major radius $R_{\rm cur}$, and vertical position $Z_{\rm cur}$, plasma energy $W$ in [MJ], central electron temperature $\tempe$ [keV], innermost LCFS major radius $R_{\rm in}$ [m], top LCFS vertical position $Z_{\rm top}$ [m], minor radius $a$ [m], outermost LCFS major radius $R_{\rm out}$, vacuum poloidal flux on the plasma boundary $\psi_{\rm ext,b}$. Color coding is the same as in the previous figure.}
		\label{figure3} 
	\end{center}
\end{figure}

In the set of \fref{figure1}, \fref{figure2}, and \fref{figure3}, the result of the benchmark on the full--discharge evolution is provided. 
As it can be seen from the three plots, the agreement between the new code FEQIS and the established SPIDER code is excellent. A small discrepancy can be seen randomly around each trace, which is expected since the codes use different algorithms. However in FEQIS a higher sensitivity of the plasma shape on details of the edge pressure gradient is observed, leading to a slightly larger plasma Shafranov shift. This results in a visible, albeit small, systematic difference in the innermost plasma boundary major radius $R_{\rm in}$ and the plasma vertical top position $Z_{\rm top}$.
\subsection{Comparison between different levels of iterations accuracy}
Before concluding, three runs have been carried out, where the results are compared when running FEQIS with full non--linear iterations at the same time step (slower mode), without non--linear iterations but calling the GSE solver at each time step, and finally when sparsely calling the GSE solver depending on the background variation of plasma parameters. 

\begin{figure}
	\begin{center}
		\includegraphics[width=150mm]{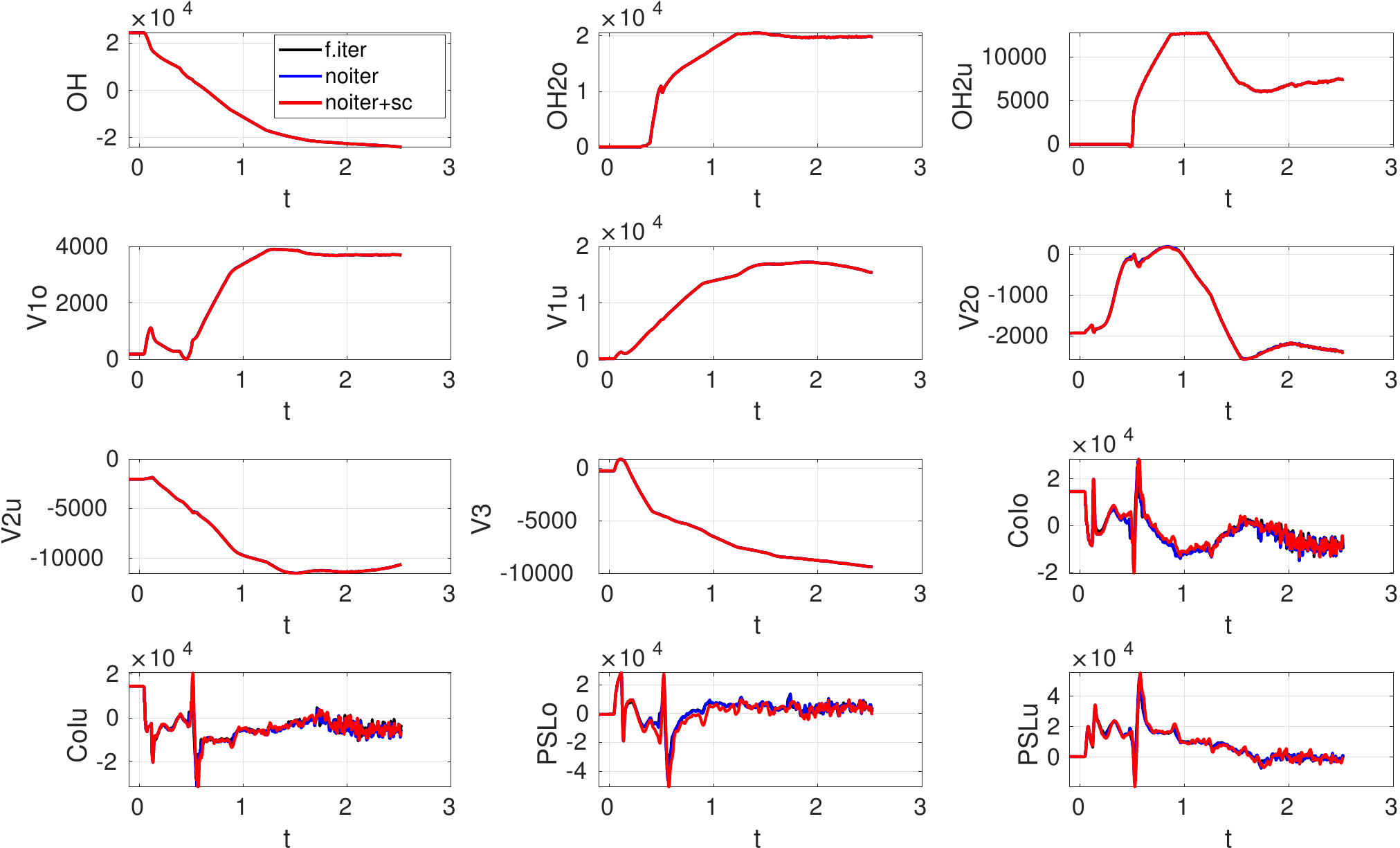}
		\caption{\small Comparison of time traces when running FEQIS with different approximations. "f.iter": full non--linear iterations. "noiter": only 1 call per time step, at all time steps. "noiter+sc": only 1 call per time step, but the GSE solver is called only if the plasma is changing in time depending on the rate of change. The time traces are all the active and passive coil currents in units of [A].}
		\label{figure41} 
	\end{center}
\end{figure}

\begin{figure}
	\begin{center}
		\includegraphics[width=150mm]{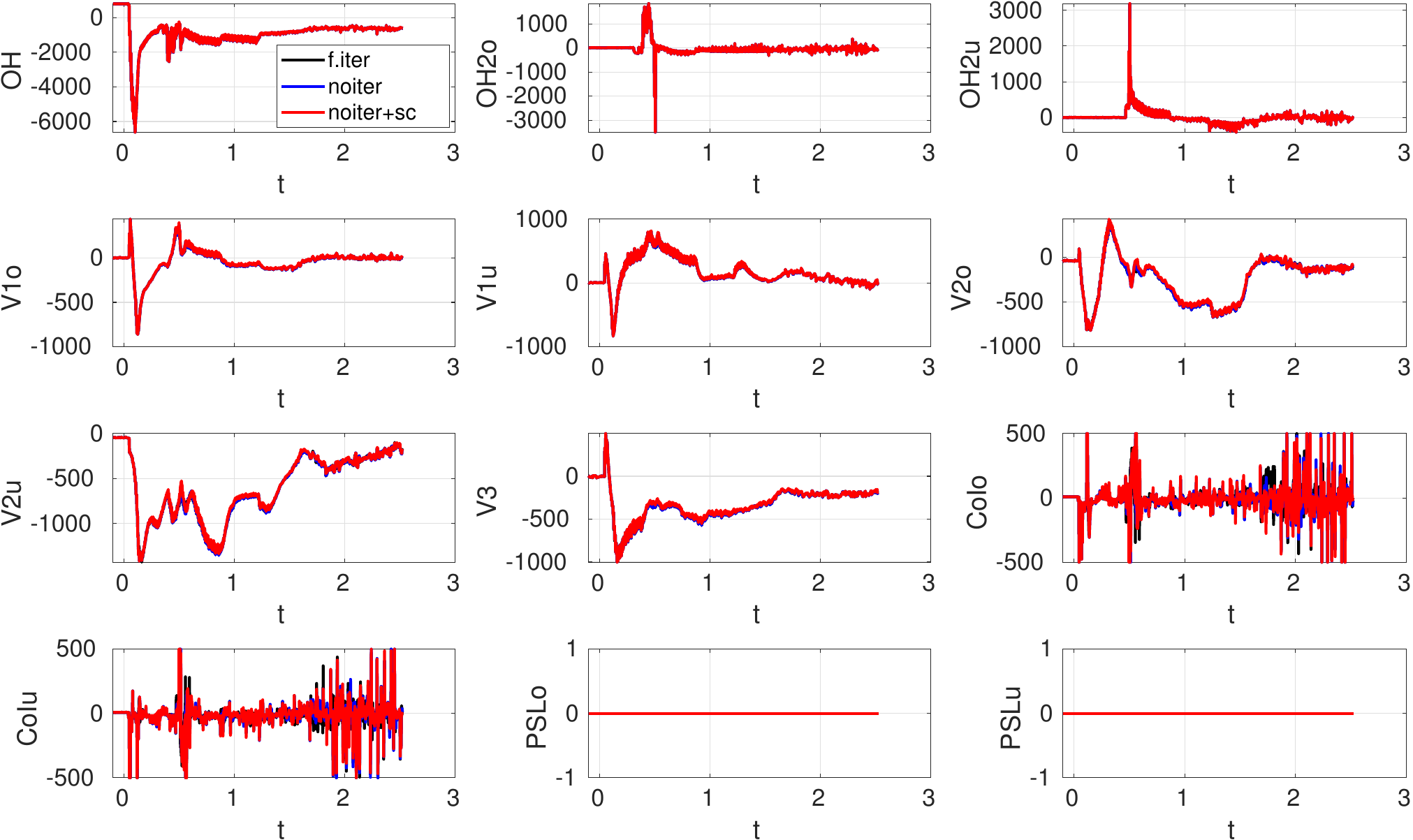}
		\caption{\small Color coding is the same as in the previous figure. Here the time traces are the voltages in [V] applied to the active coils (PSLo and PSLu are passive coils).}
		\label{figure42} 
	\end{center}
\end{figure}

\begin{figure}
	\begin{center}
		\includegraphics[width=180mm]{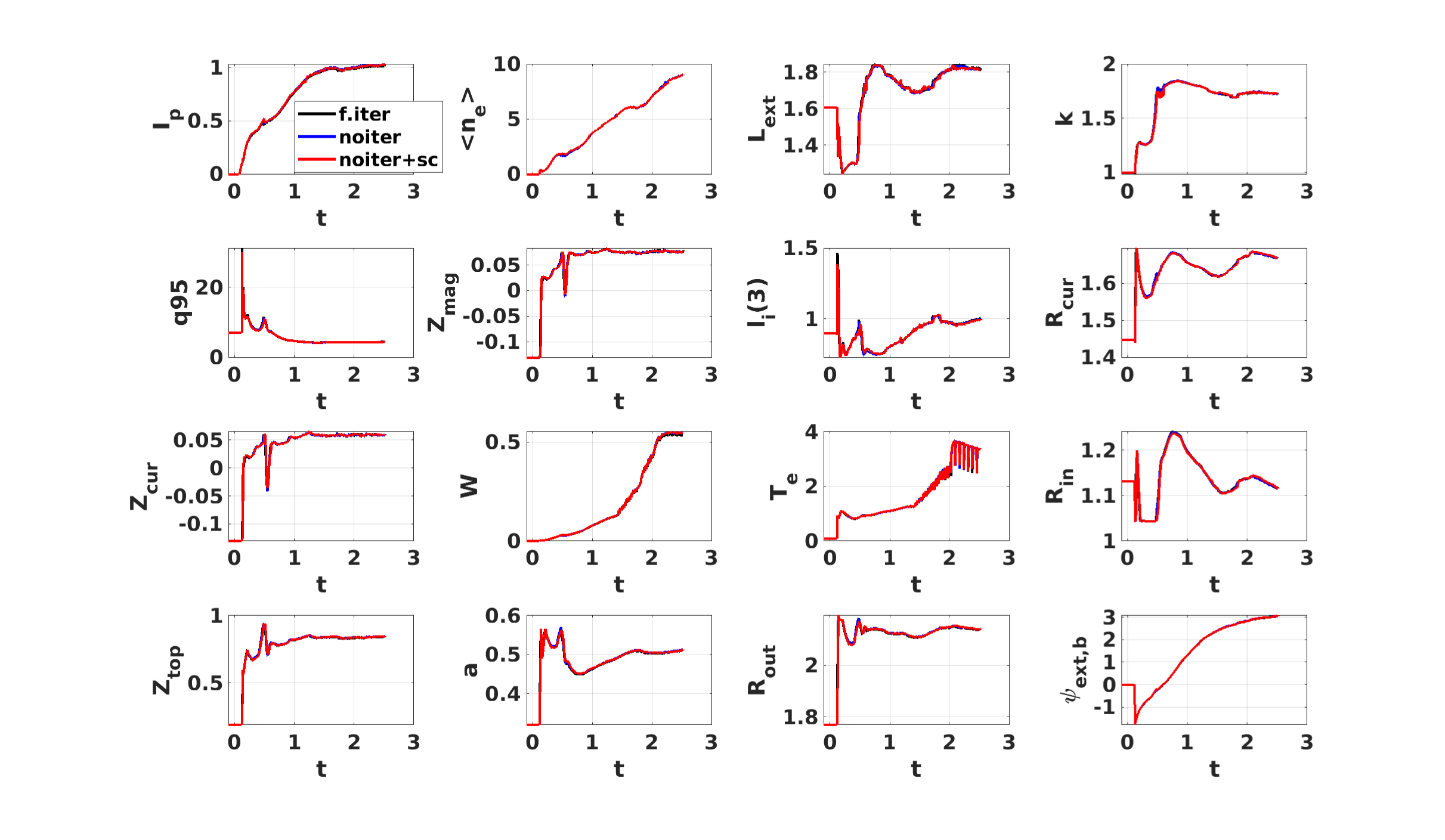}
		\caption{\small Color coding is the same as in the previous figure. Here the time traces are: plasma current $\ipl$ [MA], line averaged density $\langle \dense \rangle$ in $10^{19} m^{-3}$, plasma external inductance $L_{\rm ext}$ normalized to $2\mu_0 R$, plasma elongation $k$, $q_{95}$, magnetic axis vertical position $Z_{\rm mag}$, internal inductance $l_i(3)$, current centroid major radius $R_{\rm cur}$, and vertical position $Z_{\rm cur}$, plasma energy $W$ in [MJ], central electron temperature $\tempe$ [keV], innermost LCFS major radius $R_{\rm in}$ [m], top LCFS vertical position $Z_{\rm top}$ [m], minor radius $a$ [m], outermost LCFS major radius $R_{\rm out}$, vacuum poloidal flux on the plasma boundary $\psi_{\rm ext,b}$. Color coding is the same as in the previous figure.}
		\label{figure43} 
	\end{center}
\end{figure}

The comparison is displayed in figures (\ref{figure41}, \ref{figure42}, \ref{figure43}).

It can be seen that there is practically no difference between the various runs (the basic time step used here is 0.2 ms). However, the run with full non--linear iterations is much slower, requiring around 4 minutes to be performed for this short time window of 2.5 s of discharge, whereas the other two modes take respectively about 30 s when calling the GSE at every time step and about 20 s when calling it sparsely depending on the steadiness of the plasma.
Notice that the sparse calls could be tailored to follow fast events and be more relaxed during quiet phases, but the precise criterion is not given as part of the code.
\begin{figure}
	\begin{center}
		\includegraphics[width=115mm]{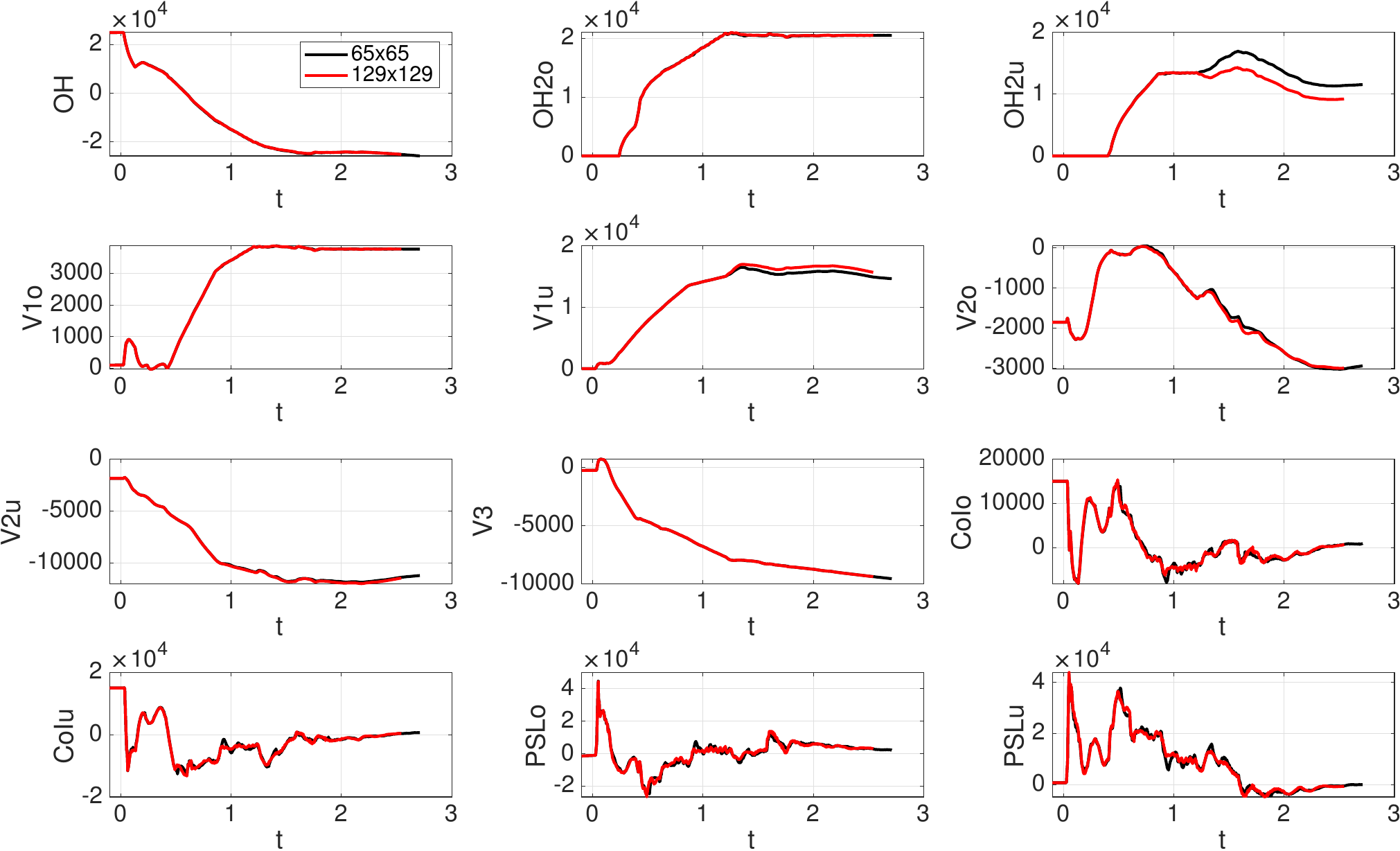}
		\includegraphics[width=115mm]{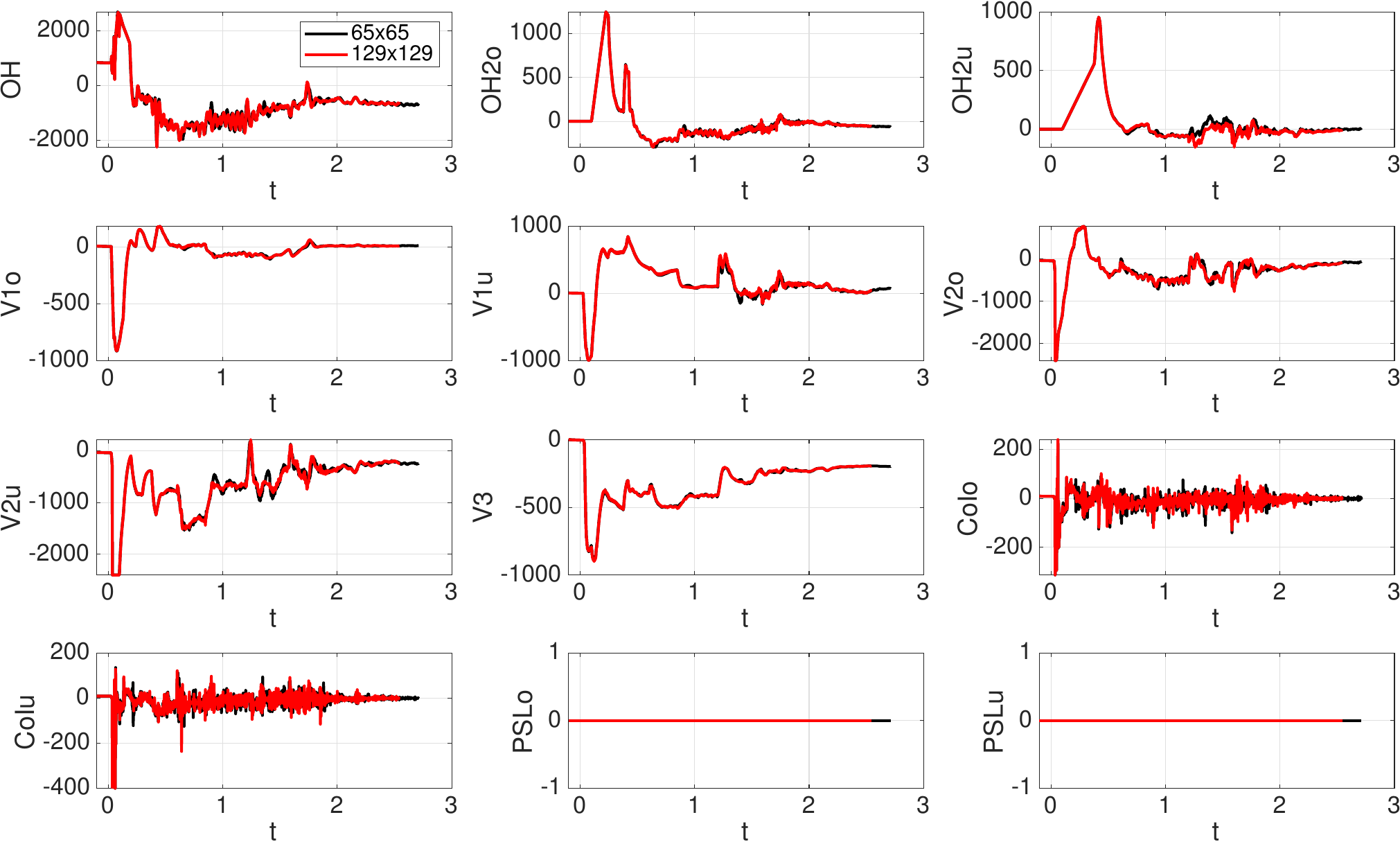}
		\includegraphics[width=115mm]{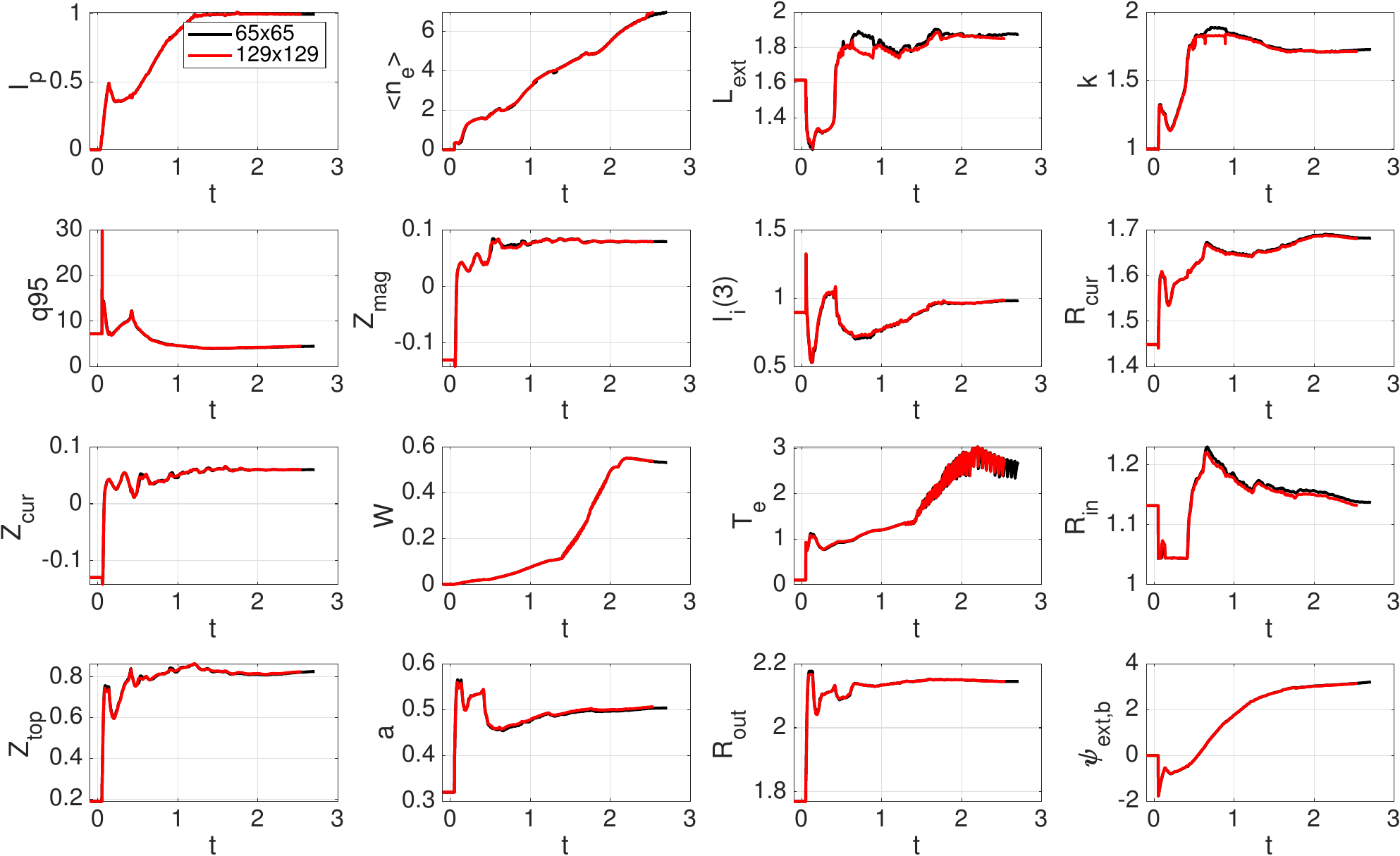}
		\caption{\small Same set as in figures (\ref{figure41},\ref{figure42},\ref{figure43}), but for a comparison using a rectangular grid of 65x65 (black lines) or of 129x129 (red lines).}
		\label{figure5} 
	\end{center}
\end{figure}

Finally, a comparison using 65x65 or 129x129 grid is presented in figure(\ref{figure5}). Notably the kinetic quantities and most of the geometric quantities do not deviate by reducing the resolution. The only exception is the coil current that is modified a bit (around $\sim 15\%$) is the OH2u coil current. This coil is used for strike--point control, as such it is more sensitive to the grid resolution which impacts the accuracy with which the separatrix legs and their intersection points with the divertor targets are obtained.

\section{Conclusions}
In this work, a new dynamical Grad--Shafranov and circuit equation solver FEQIS is presented. This code primary scope is to do forward modeling of the plasma equilibrium evolution inside predictive modeling and particularly inside the flight simulator framework.
Its various modes of operation and details of the algorithms employed to solve specific aspects of the free--boundary problem are presented.

A comparison with the SPIDER GSE solver is carried out inside the flight simulator Fenix framework, running a full--discharge prediction for an existing H--mode from the ASDEX Upgrade.
The agreement between FEQIS and SPIDER is excellent and poses the basis for further developments of FEQIS.
In particular, pushing the speed of the code will be the main direction going forward, whereas as of now FEQIS and SPIDER have similar computational speed when both are run with rectangular grid description.
Moreover, ferro--magnetic materials will be included via the magnetization currents method.


\section*{Acknowledgments}
\addcontentsline{toc}{section}{Acknowledgments}

This work has strongly benefited from discussions with various Colleagues who are experts in this field: Dr. O. Maj (IPP Garching); Dr. M. Mattei (ENEA, CREATE Consortium, University of Naples).

\newpage


\bibliographystyle{aomalpha}
\bibliography{references_list.bib}

\providecommand{\etalchar}[1]{$^{#1}$}
\providecommand{\bysame}{\leavevmode\hbox to3em{\hrulefill}\thinspace}
\providecommand{\noopsort}[1]{}
\providecommand{\mr}[1]{\href{http://www.ams.org/mathscinet-getitem?mr=#1}{MR~#1}}
\providecommand{\zbl}[1]{\href{http://www.zentralblatt-math.org/zmath/en/search/?q=an:#1}{Zbl~#1}}
\providecommand{\jfm}[1]{\href{http://www.emis.de/cgi-bin/JFM-item?#1}{JFM~#1}}
\providecommand{\arxiv}[1]{\href{http://www.arxiv.org/abs/#1}{arXiv~#1}}
\providecommand{\doi}[1]{\url{https://doi.org/#1}}
\providecommand{\MR}{\relax\ifhmode\unskip\space\fi MR }
\providecommand{\MRhref}[2]{%
  \href{http://www.ams.org/mathscinet-getitem?mr=#1}{#2}
}
\providecommand{\href}[2]{#2}
\begin{thebibliography}{RPFtAUT16}

\bibitem[AAM15]{createeqcodepaper1}
\bgroup\scshape{}R.~Albanese\egroup{}, \bgroup\scshape{}R.~Ambrosino\egroup{},
  and \bgroup\scshape{}M.~Mattei\egroup{}, {CREATE-NL+}: A robust
  control-oriented free boundary dynamic plasma equilibrium solver,
  \emph{Fusion Engineering and Design} \textbf{96-97} (2015), 664--667.
  \doi{10.1016/j.fusengdes.2015.06.162}.

\bibitem[BL77]{lacknerseginductanceformula}
\bgroup\scshape{}G.~Becker\egroup{} and \bgroup\scshape{}K.~Lackner\egroup{},
  Free-boundary equilibrium computations for strongly elongated {Belt} {Pinch}
  {Plasmas},  \emph{Nucl. Fusion} \textbf{17} (1977), 903.

\bibitem[BLF84]{blumeqreview}
\bgroup\scshape{}J.~Blum\egroup{} and \bgroup\scshape{}J.~Le~Foll\egroup{},
  Plasma equilibrium evolution at the resistive diffusion timescale,
  \emph{Computer Physics Reports} \textbf{1} no.~7 (1984), 465--494.
  \doi{10.1016/0167-7977(84)90013-3}.

\bibitem[DFB{\etalchar{+}}22]{nngssolvernaturedegravepaper4}
\bgroup\scshape{}J.~Degrave\egroup{}, \bgroup\scshape{}F.~Felici\egroup{},
  \bgroup\scshape{}J.~Buchli\egroup{}, \bgroup\scshape{}M.~Neunert\egroup{},
  \bgroup\scshape{}B.~Tracey\egroup{}, \bgroup\scshape{}F.~Carpanese\egroup{},
  \bgroup\scshape{}T.~Ewalds\egroup{}, \bgroup\scshape{}R.~Hafner\egroup{},
  \bgroup\scshape{}A.~Abdolmaleki\egroup{}, \bgroup\scshape{}D.~de~las
  Casas\egroup{}, and \bgroup\scshape{}others\egroup{}, Magnetic control of
  tokamak plasmas through deep reinforcement learning,  \emph{Nature}
  \textbf{602} (2022), 414--419. \doi{10.1038/s41586-021-04301-9}.

\bibitem[ELSV24]{mmontecarlogssolverpaper1}
\bgroup\scshape{}H.~C. Elman\egroup{}, \bgroup\scshape{}J.~Liang\egroup{}, and
  \bgroup\scshape{}T.~Sánchez-Vizuet\egroup{}, Multilevel {Monte Carlo}
  methods for the {Grad-Shafranov} free boundary problem,  \emph{Computer
  Physics Communications} \textbf{298} (2024), 109099.
  \doi{10.1016/j.cpc.2024.109099}.

\bibitem[FAC{\etalchar{+}}13]{astracode2}
\bgroup\scshape{}E.~Fable\egroup{}, \bgroup\scshape{}C.~Angioni\egroup{},
  \bgroup\scshape{}F.~J. Casson\egroup{}, \bgroup\scshape{}D.~Told\egroup{},
  \bgroup\scshape{}A.~A. Ivanov\egroup{}, \bgroup\scshape{}F.~Jenko\egroup{},
  \bgroup\scshape{}R.~M. McDermott\egroup{}, \bgroup\scshape{}S.~Y.
  Medvedev\egroup{}, \bgroup\scshape{}G.~V. Pereverzev\egroup{},
  \bgroup\scshape{}F.~Ryter\egroup{}, and \bgroup\scshape{}others\egroup{},
  Novel free-boundary equilibrium and transport solver with theory-based models
  and its validation against {ASDEX} upgrade current ramp scenarios,
  \emph{Plasma Phys. Contr. Fusion} \textbf{55} (2013), 124028.
  \doi{10.1088/0741-3335/55/12/124028}.

\bibitem[FJT{\etalchar{+}}22]{fenixfableref}
\bgroup\scshape{}E.~Fable\egroup{}, \bgroup\scshape{}F.~Janky\egroup{},
  \bgroup\scshape{}W.~Treutterer\egroup{},
  \bgroup\scshape{}M.~Englberger\egroup{},
  \bgroup\scshape{}R.~Schramm\egroup{}, \bgroup\scshape{}M.~Muraca\egroup{},
  \bgroup\scshape{}C.~Angioni\egroup{}, \bgroup\scshape{}O.~Kudlacek\egroup{},
  \bgroup\scshape{}E.~Poli\egroup{}, \bgroup\scshape{}M.~Reich\egroup{}, and
  \bgroup\scshape{}others\egroup{}, The modeling of a tokamak plasma discharge,
  from first principles to a flight simulator,  \emph{Plasma Phys. Contr.
  Fusion} \textbf{64} (2022), 044002. \doi{10.1088/1361-6587/ac466b}.

\bibitem[Fau20]{nicepaperref}
\bgroup\scshape{}B.~Faugeras\egroup{}, An overview of the numerical methods for
  tokamak plasma equilibrium computation implemented in the {NICE} code,
  \emph{Fus. Eng. Design} \textbf{160} (2020), 112020.
  \doi{10.1016/j.fusengdes.2020.112020}.

\bibitem[FMS92]{fbeextrapcurrentspasche}
\bgroup\scshape{}B.~Fornberg\egroup{} and
  \bgroup\scshape{}R.~Meyer-Spasche\egroup{}, A finite difference procedure for
  a class of free boundary problems,  \emph{Journal of Computational Physics}
  \textbf{102} no.~1 (1992), 72--77. \doi{10.1016/S0021-9991(05)80006-3}.

\bibitem[GL04]{contdynamicsgssolverpaper6}
\bgroup\scshape{}P.~A. Gourdain\egroup{} and \bgroup\scshape{}J.~N.
  Leboeuf\egroup{}, Contour dynamics method for solving the {Grad–Shafranov}
  equation with applications to high beta equilibria,  \emph{Physics of
  Plasmas} \textbf{11} no.~9 (2004), 4372--4381. \doi{10.1063/1.1776174}.

\bibitem[HSB{\etalchar{+}}24]{tokamakergscodepaper}
\bgroup\scshape{}C.~Hansen\egroup{}, \bgroup\scshape{}I.~Stewart\egroup{},
  \bgroup\scshape{}D.~Burgess\egroup{}, \bgroup\scshape{}M.~Pharr\egroup{},
  \bgroup\scshape{}S.~Guizzo\egroup{}, \bgroup\scshape{}F.~Logak\egroup{},
  \bgroup\scshape{}A.~Nelson\egroup{}, and
  \bgroup\scshape{}C.~Paz-Soldan\egroup{}, {TokaMaker}: An open-source
  time-dependent {Grad-Shafranov} tool for the design and modeling of
  axisymmetric fusion devices,  \emph{Computer Physics Communications}
  \textbf{298} (2024), 109111. \doi{10.1016/j.cpc.2024.109111}.

\bibitem[HBB{\etalchar{+}}15]{cedrescodepaper}
\bgroup\scshape{}H.~Heumann\egroup{}, \bgroup\scshape{}J.~Blum\egroup{},
  \bgroup\scshape{}C.~Boulbe\egroup{}, \bgroup\scshape{}B.~Faugeras\egroup{},
  \bgroup\scshape{}G.~Selig\egroup{}, \bgroup\scshape{}J.-M. Ané\egroup{},
  \bgroup\scshape{}S.~Brémond\egroup{},
  \bgroup\scshape{}V.~Grandgirard\egroup{},
  \bgroup\scshape{}P.~Hertout\egroup{}, and
  \bgroup\scshape{}E.~Nardon\egroup{}, Quasi-static free-boundary equilibrium
  of toroidal plasma with {CEDRES++}: Computational methods and applications,
  \emph{Journal of Plasma Physics} \textbf{81} no.~3 (2015), 905810301.
  \doi{10.1017/S0022377814001251}.

\bibitem[HS14]{specelemgssolverpaper5}
\bgroup\scshape{}E.~C. Howell\egroup{} and \bgroup\scshape{}C.~R.
  Sovinec\egroup{}, Solving the {Grad–Shafranov} equation with spectral
  elements,  \emph{Computer Physics Communications} \textbf{185} no.~5 (2014),
  1415--1421. \doi{10.1016/j.cpc.2014.02.008}.

\bibitem[I{\etalchar{+}}05]{spidercoderef}
\bgroup\scshape{}A.~A. Ivanov\egroup{} and \bgroup\scshape{}others\egroup{},
  \emph{32nd EPS Conf. on Plasma Physics} \textbf{29C} (2005), 5.063.

\bibitem[IKMP09]{spidercoderef2}
\bgroup\scshape{}A.~A. Ivanov\egroup{},
  \bgroup\scshape{}R.~Khayrutdinov\egroup{},
  \bgroup\scshape{}S.~Medvedev\egroup{}, and
  \bgroup\scshape{}Y.~Poshekhonov\egroup{}, The {SPIDER} code - solution of
  direct and inverse problems for free boundary tokamak plasma equilibrium,
  \emph{Keldysh Institute preprints} \textbf{39} (2009), 24.

\bibitem[JKG{\etalchar{+}}24]{nngssolverpinnspaper1}
\bgroup\scshape{}B.~Jang\egroup{}, \bgroup\scshape{}A.~A. Kaptanoglu\egroup{},
  \bgroup\scshape{}R.~Gaur\egroup{}, \bgroup\scshape{}S.~Pan\egroup{},
  \bgroup\scshape{}M.~Landreman\egroup{}, and
  \bgroup\scshape{}W.~Dorland\egroup{}, {Grad–Shafranov} equilibria via
  data-free physics informed neural networks,  \emph{Physics of Plasmas}
  \textbf{31} no.~3 (2024), 032510. \doi{10.1063/5.0188634}.

\bibitem[JFET21]{fenixjanky1}
\bgroup\scshape{}F.~Janky\egroup{}, \bgroup\scshape{}E.~Fable\egroup{},
  \bgroup\scshape{}M.~Englberger\egroup{}, and
  \bgroup\scshape{}W.~Treutterer\egroup{}, Validation of the {Fenix} {ASDEX}
  {Upgrade} flight simulator,  \emph{Fusion Engineering and Design}
  \textbf{163} (2021), 112126. \doi{10.1016/j.fusengdes.2020.112126}.

\bibitem[Jeo15]{TEScodepaper}
\bgroup\scshape{}Y.~M. Jeon\egroup{}, Development of a free boundary tokamak
  equilibrium solver ({TES}) for advanced study of tokamak equilibria,
  \emph{Journal of the Korean Physical Society} \textbf{67} (2015), 843--853.
  \doi{10.3938/jkps.67.843}.

\bibitem[KL93]{dinarefpaper}
\bgroup\scshape{}R.~R. Khayrutdinov\egroup{} and \bgroup\scshape{}V.~E.
  Lukash\egroup{}, Studies of plasma equilibrium and transport in a tokamak
  fusion device with the inverse-variable technique,  \emph{Journal of
  Computational Physics} \textbf{109} no.~2 (1993), 193--201.
  \doi{10.1006/jcph.1993.1211}.

\bibitem[Lac76]{lacknereqcodepaper}
\bgroup\scshape{}K.~Lackner\egroup{}, Computation of ideal {MHD} equilibria,
  \emph{Computer Physics Communications} \textbf{12} no.~1 (1976), 33--44.
  \doi{10.1016/0010-4655(76)90008-4}.

\bibitem[MAA{\etalchar{+}}24]{nngssolvergreenpaper2}
\bgroup\scshape{}J.~McClenaghan\egroup{}, \bgroup\scshape{}C.~Akçay\egroup{},
  \bgroup\scshape{}T.~B. Amara\egroup{}, \bgroup\scshape{}X.~Sun\egroup{},
  \bgroup\scshape{}S.~Madireddy\egroup{}, \bgroup\scshape{}L.~L. Lao\egroup{},
  \bgroup\scshape{}S.~E. Kruger\egroup{}, and \bgroup\scshape{}O.~M.
  Meneghini\egroup{}, Augmenting machine learning of {Grad–Shafranov}
  equilibrium reconstruction with {Green}'s functions,  \emph{Physics of
  Plasmas} \textbf{31} no.~8 (2024), 082507. \doi{10.1063/5.0213625}.

\bibitem[MFA{\etalchar{+}}23]{fenixmodelsmuraca}
\bgroup\scshape{}M.~Muraca\egroup{}, \bgroup\scshape{}E.~Fable\egroup{},
  \bgroup\scshape{}C.~Angioni\egroup{}, \bgroup\scshape{}T.~Luda\egroup{},
  \bgroup\scshape{}P.~David\egroup{}, \bgroup\scshape{}H.~Zohm\egroup{},
  \bgroup\scshape{}A.~Di~Siena\egroup{}, and \bgroup\scshape{}the ASDEX
  Upgrade~Team\egroup{}, Reduced transport models for a tokamak flight
  simulator,  \emph{Plasma Phys. Contr. Fusion} \textbf{65} (2023), 035007.
  \doi{10.1088/1361-6587/acb2c6}.

\bibitem[PKF16]{mimeticsegssolverpaper3}
\bgroup\scshape{}A.~Palha\egroup{}, \bgroup\scshape{}B.~Koren\egroup{}, and
  \bgroup\scshape{}F.~Felici\egroup{}, A mimetic spectral element solver for
  the {Grad–Shafranov} equation,  \emph{Journal of Computational Physics}
  \textbf{316} (2016), 63--93. \doi{10.1016/j.jcp.2016.04.002}.

\bibitem[PCF{\etalchar{+}}13]{fasthighordergssolver2}
\bgroup\scshape{}A.~Pataki\egroup{}, \bgroup\scshape{}A.~J. Cerfon\egroup{},
  \bgroup\scshape{}J.~P. Freidberg\egroup{},
  \bgroup\scshape{}L.~Greengard\egroup{}, and
  \bgroup\scshape{}M.~O’Neil\egroup{}, A fast, high-order solver for the
  {Grad–Shafranov} equation,  \emph{Journal of Computational Physics}
  \textbf{243} (2013), 28--45. \doi{10.1016/j.jcp.2013.02.045}.

\bibitem[PY91]{astracode1}
\bgroup\scshape{}G.~V. Pereverzev\egroup{} and \bgroup\scshape{}P.~N.
  Yushmanov\egroup{},  \emph{IPP Report} \textbf{5/42} (1991).

\bibitem[RPFtAUT16]{sintransformeqgarching}
\bgroup\scshape{}M.~Rampp\egroup{}, \bgroup\scshape{}R.~Preuss\egroup{},
  \bgroup\scshape{}R.~Fischer\egroup{}, and \bgroup\scshape{}the ASDEX
  Upgrade~Team\egroup{}, {GPEC}: A real-time–capable tokamak equilibrium
  code,  \emph{Fusion Science and Technology} \textbf{70} no.~1 (2016), 1--13.
  \doi{10.13182/FST15-154}.

\bibitem[RCRF16]{accuratederivgssolverpaper4}
\bgroup\scshape{}L.~Ricketson\egroup{}, \bgroup\scshape{}A.~Cerfon\egroup{},
  \bgroup\scshape{}M.~Rachh\egroup{}, and
  \bgroup\scshape{}J.~Freidberg\egroup{}, Accurate derivative evaluation for
  any {Grad–Shafranov} solver,  \emph{Journal of Computational Physics}
  \textbf{305} (2016), 744--757. \doi{10.1016/j.jcp.2015.11.015}.

\bibitem[Sha60]{gradshafranoveq}
\bgroup\scshape{}V.~D. Shafranov\egroup{}, Equilibrium of a plasma toroid in a
  magnetic field,  \emph{Sov. Phys. - JETP} \textbf{10} (1960), 775.

\bibitem[WSRS24]{nngssolverfbenetpaper3}
\bgroup\scshape{}Z.~Wang\egroup{}, \bgroup\scshape{}X.~Song\egroup{},
  \bgroup\scshape{}T.~Rafiq\egroup{}, and
  \bgroup\scshape{}E.~Schuster\egroup{}, Neural-network-based free-boundary
  equilibrium solver to enable fast scenario simulations,  \emph{IEEE
  Transactions on Plasma Science} \textbf{52} no.~9 (2024), 4147--4153.
  \doi{10.1109/TPS.2024.3375284}.

\bibitem[WDF{\etalchar{+}}24]{fenixinterfref}
\bgroup\scshape{}C.~Wu\egroup{}, \bgroup\scshape{}P.~David\egroup{},
  \bgroup\scshape{}E.~Fable\egroup{}, \bgroup\scshape{}D.~Frattolillo\egroup{},
  \bgroup\scshape{}L.~E. Di~Grazia\egroup{},
  \bgroup\scshape{}M.~Mattei\egroup{}, \bgroup\scshape{}M.~Siccinio\egroup{},
  \bgroup\scshape{}W.~Treutterer\egroup{}, and
  \bgroup\scshape{}H.~Zohm\egroup{}, Architecture design and internal
  implementation of a universal coupling between controllers and physics in a
  tokamak flight simulator,  \emph{Fusion Science and Technology} \textbf{80}
  no.~6 (2024), 766--771. \doi{10.1080/15361055.2023.2234741}.

\end{thebibliography}
\addcontentsline{toc}{section}{References}

\end{document}